\documentclass[final,5p,times,onecolumn,unknownkeysallowed]{elsarticle}

\usepackage{doi}
\usepackage{hyperref}
\hypersetup{
	colorlinks=true,        
	linkcolor=blue,         
	citecolor=cyan,         
}

\usepackage[utf8x]{inputenc}
\usepackage{graphicx}
\usepackage{amssymb}
\usepackage{amsmath}
\usepackage{amsthm}
\usepackage{lineno}
\usepackage{hyperref}
\usepackage{multirow}
\usepackage{bigints}

\usepackage{dcolumn}
\usepackage{bm}
\usepackage{color}
\usepackage{xcolor}


\usepackage{dcolumn}
\usepackage{bm}
\usepackage{color}
\usepackage{cuted}

\hypersetup{colorlinks=true,linkcolor=cyan,citecolor=cyan,urlcolor=blue,bookmarks=true}\modulolinenumbers[200]

\biboptions{comma,sort&compress}

\journal{Journal Name}

\begin{document}

\begin{frontmatter}
    \title{New analytical model of rotating black hole with dark matter halo: constraints from EHT observations and accretion disk}
    
\author[mainaddress0,mainaddress6,mainaddress4]
{Uktamjon Uktamov\corref{cor1}}
\cortext[cor1]{Corresponding author}
\ead{uktam.uktamov11@gmail.com}%
\author[mainaddress1,mainaddress2,mainaddress3]
{Sanjar Shaymatov}
\ead{sanjar@astrin.uz}
\author[mainaddress0,mainaddress4,mainaddress5,]
{Bobomurat Ahmedov}
\ead{ahmedov@astrin.uz}
\author[mainaddress0]{Chengxun Yuan}
\ead{yuancx@hit.edu.cn}
\address[mainaddress0]{School of Physics, Harbin Institute of Technology, Harbin 150001, People’s Republic of China}

\address[mainaddress1]{Institute of Fundamental and Applied Research, National Research University TIIAME, Kori Niyoziy 39, Tashkent 100000, Uzbekistan}
\address[mainaddress2]{Institute for Theoretical Physics and Cosmology,
Zhejiang University of Technology, Hangzhou 310023, China}
\address[mainaddress4]{Institute for Advanced Studies, New Uzbekistan University, Movarounnahr str. 1, Tashkent 100000, Uzbekistan}
\address[mainaddress5]{Institute of Theoretical Physics, National University of Uzbekistan, Tashkent 100174, Uzbekistan}
\address[mainaddress6]{Tashkent University of Applied Sciences, Gavhar Str. 1, Tashkent 100149, Uzbekistan}
\address[mainaddress3]{Tashkent State Technical University, Tashkent 100095, Uzbekistan}

\date{Received: date / Accepted: date}

\begin{abstract}
In this paper, we start from a static black hole (BH) immersed in a Dehnen-type dark matter (DM) halo and employ the Newman-Janis algorithm (NJA) to generate the rotating black hole solution with a dark matter halo. Also, we have checked the validity of the obtained space-time. Then we study optical properties of newly obtained rotating BH in DM halo, including the shadow’s geometrical shape, deflection angle of light based Ono, Ishihara and Asada (OID) method, photon sphere and the dependence of the shadow radius on DM parameters.   Additionally, assuming that spacetime of a supermassive black hole (SMBH) is described by the newly obtained rotating BH solution, we analyze the parameters of the model with shadow size estimates based on the  Event Horizon Telescope (EHT) and Gravity collaboration observations of M87$^*$ and Sgr A$^*$ SMBHs. Then we have used Markov Chain Monte Carlo (MCMC) analysis to constrain DM parameters $\rho_s$, $r_s$ and BH mass $M$, BH spin $a$, also we show that best-fit values for the parameters $\rho_s$, $r_s$ are well agreement with  previous results which indicate physically reasonability of the our model. Finally, we have analyzed the electromagnetic radiation flux of the rotating BH in the DM halo employing a ray tracing code. 
\end{abstract}


\end{frontmatter}
\linenumbers

\section{Introduction}
One of the most fascinating predictions of General Relativity (GR) was the formation of black holes (BH) at the end state of evolution of massive stars and the assumption of existence of black holes in the center of galaxies \cite{LIGOScientific:2016aoc,2016PhRvL.116x1102A} which are responsible for high energetics of active galactic nuclei (AGN). The last decade was related to the triumphal revolutionary discoveries in relativistic astrophysics. However, these essential observations were not enough to explain the nature of the interaction between black holes and their surroundings. Therefore, understanding a behavior of the dark matter (DM) is crucial task in GR as there are enough evidence that BHs are surrounded by DM halo \cite{2015NatPh..11..245I,2018Natur.562...51B}. Although, numerous efforts have been done to understand the nature of the DM halo, there is still only one way, gravitational interactions, to understand DM. Thus, finding exact analytical model for DM halo is utmost problem in GR. There are several analytical models, Einasto \cite{Graham:2006ae}, Navarro-Frenk-White \cite{1996ApJ...462..563N}, Burkert \cite{1995ApJ...447L..25B} and the Dehnen model \cite{1993MNRAS.265..250D,2021A&A...654A..53S,Al-Badawi:2024asn,2025PhRvD.111d4060M,2025PhLB..86239300S,2024PhLB..85538797S} to describe the interaction between BH and DM. In this work, we have chosen core model of the Dehnen-type dark matter halo with the known exact analytical solution for static BH in DM halo and then using Newman-Janis algorithm we generated the novel rotating black hole solution with a dark matter halo.

Recent studies within the Dehnen-type DM halo framework have also examined interactions between DM and BHs from various viewpoints. In particular, these studies thoroughly examine black holes embedded within Dehnen-type dark matter halos exploring the influence of the DM halo on quasinormal modes, the BH shadow, and the photon sphere radius \cite{Xamidov:2025prl}, alongside analyzing gravitational waveforms generated by periodic orbits \cite{2025arXiv250405236A} for the BH-Dehnen-type DM halo solution. Furthermore, the null geodesics and the thermodynamics of the effective BH-DM halo system wer studied in~\cite{2024PDU....4601683G}, which is followed by an analysis that constrains the parameters of the DM halo \cite{2025PDU....4701805X}.

Observations from the EHT, specifically images of supermassive black holes (SMBHs) Sgr A$^*$ and M87$^*$, give opportunities to compare our theoretical findings with the real observational data \cite{PhysRevD.100.024020,Bambi:2019tjh,2023CQGra..40p5007V}. In 2019, the EHT collaboration made a groundbreaking announcement by providing the first image of a black hole, a shadow-like representation of the supermassive black hole located in the center of the M87 galaxy\cite{Akiyama19L1}. Following,  researchers from the EHT project unveiled an image of the black hole at the center of the Milky Way galaxy, known as Sgr~A*, in 2022. In accordance with GR, the observed images of the two black holes, M87$^*$ and Sgr A$^*$, are consistent with the properties predicted for a Kerr black hole. This agreement indicates that these black holes are effectively characterized by the Kerr metric, which serves as the solution to the Einstein field equations for a rotating black hole\cite{2022ApJ...930L..12E}. Although Kerr-like black holes emerging from modified gravity theories or including a DM halo are not fully validated due to the relative deviation in quadrupole moments and the current measurement uncertainties in spin or angular momentum, they cannot be completely excluded \cite{2019LRR....22....4C}. In recent years, the study of black hole shadows has attracted significant attention, particularly in the context of modified gravity theories \cite{2022IJMPD..3150058Z}. Additionally, the exploration of spinning black holes has offered new insights into the optical properties and shadow features of these celestial objects \cite{2022EPJC...82..948Z}. Moreover, observational data, including those gathered by the EHT, have been crucial in constraining the range of theoretical parameters\cite{2024JCAP...05..047R}. Consequently, one of the key challenges in modern astrophysics is evaluating the nature and existence of dark matter models using the data supplied by the EHT Collaboration. In this paper, we obtain constraints for the characteristic density $\rho_s$ and the characteristic scale factor $r_s$ of the DM halo using the observed data mentioned above. 

The paper organized as follows: in Sec.~\ref{Sec:II}, we generate a solution for a rotating BH surrounded by a Dehnen-type dark matter halo using Newman-Janis algorithm. In addition, event horizon of the rotating BH with a dark matter halo is analyzed in this section. The following null geodesics equations are derived using the Hamilton-Jacobi equation, the photon sphere around a rotating BH in a dark matter halo is analyzed, and finally, the shadow of the rotating BH in dark matter halo is investigated in Sec.~\ref{Sec.III}. Subsequently, deflection angle of light by a rotating BH in DM halo and effect of the DM halo on deflection angle $\Hat{\alpha}$ is investigated in Sec.\ref{Sec.4}. Then we obtain constraints for the BH parameters $r_s$ and $\rho_s$ applying the EHT and Gravity collaboration observation data to the obtained theoretical results in Sec.~\ref{sec.EHT} and then we employed MCMC analysis to find best-fit values of the DM parameters $\rho_s$, $r_s$. Also we obtain image of the thin accretion in rotating BH surrounded by Dehnen-type DM halo in Sec. ~\ref{Sec.VI}. Finally, we give our main conclusions in Sec.~\ref{Sec.conclusion}.

\section{Spacetime of rotating black hole with the dark matter halo}\label{Sec:II}

 The authors in \cite{UktamjonUktamov:2025emm} have found a novel static black hole solution surrounded by Dehnen-type dark matter halo in the form:
\begin{eqnarray}\label{eq.full-line}
    ds^2=-f(r)dt^2+\frac{1}{f(r)}dr^2+r^{2}\left(
d\theta ^{2}+\sin ^{2}\theta d\phi ^{2}\right)\, , 
\end{eqnarray}
where 
\begin{eqnarray}
    f(r)=
    1-\frac{2M}{r}-8\pi\rho_sr_s^2\log{\left(1+\frac{r_s}{r}\right)}\ , 
\end{eqnarray}
 and $\rho_s$ and $r_s$ are the characteristic density and characteristic scale factor of the DM halo.

Now using Newman and Janis method (NJA) \cite{2017EPJC...77..542T,2014PhRvD..90f4041A,Azreg-Ainou:2014aqa,2014PhLB..730...95A,Newman} we will generate rotating BH space-time in dark matter halo. To do it, we start with coordinate transformation from the Boyer-Lindquist (BL) coordinates $(t, r, \theta, \phi)$ to the Eddington-Finkelstein (EF) coordinates $(u, r, \theta, \phi)$:
\begin{eqnarray}\label{eq.coor. transf}
    du=dt-\frac{dr}{f(r)},
\end{eqnarray}
which gives us:
\begin{eqnarray}\label{eq.coor.transf.2}
    ds^2=-f(r)du^2-2dudr+r^2d\Omega^2,
\end{eqnarray}
with $d\Omega=d\theta^2+\sin^2{\theta}d\phi^2$.

The contravariant components of the metric tensor in null EF coordinates can be written as:
\begin{eqnarray}\label{eq.g}
    g^{\mu\nu}=-l^\mu n^\nu-l^\nu n^\mu+m^\mu\Bar{m}^\nu+m^\nu\Bar{m}^\mu,
\end{eqnarray}
where null tetrads can be expressed after performing complex coordinate transformations $u\to u-ia\cos{\theta}$, $r\to r-ia\cos{\theta}$, and resulting $f(r)\to F(r,a,\theta)$, $r^2\to\Sigma(r,a,\theta)$: 
\begin{subequations}\label{eq.null tetrads}
\begin{align}
 & l^\mu=\delta_r^\mu \\
 & m^\mu=\frac{1}{\sqrt{2\Sigma}}\left(\delta_\theta^\mu+ia\sin{\theta}(\delta_u^\mu-\delta_r^\mu)+\frac{i}{\sin{\theta}}\delta_\phi^\mu\right),
 \\
 & n^\mu=\delta_u^\mu-\frac{1}{2}F\delta_r^\mu,
 \\
 & \Bar{m}^\mu=\frac{1}{\sqrt{2\Sigma}}\left(\delta_\theta^\mu-ia\sin{\theta}(\delta_u^\mu-\delta_r^\mu)-\frac{i}{\sin{\theta}}\right).
\end{align}
\end{subequations}
Finally, we can express a rotating black hole spacetime  surrounded by Dehnen-type dark matter halo:
\begin{eqnarray}\label{eq.rotating line el.}
ds^2=g_{tt}dt^2+2g_{t\phi}dtd\phi+g_{\phi\phi}d\phi^2+g_{rr}dr^2+g_{\theta\theta}d\theta^2,
\end{eqnarray}
with metric components:
\begin{subequations}\label{eq.metric components}
\begin{eqnarray}
 g_{tt}&=&-\left[1-\frac{2Mr+2M_{D}r\log{(1+\frac{r_s}{r})^{\left(1+\frac{r}{r_s}\right)}}}{\Sigma}\right], \\ 
 g_{rr}&=&\frac{\Sigma}{\Delta},\\
  g_{t\phi}&=&-a\sin^2{\theta}\left[\frac{2Mr+2M_{D}r\log\left(1+\frac{r_s}{r}\right)^{\left(1+\frac{r}{r_s}\right)}}{\Sigma}\right],\\
g_{\phi\phi}&=&\sin^4{\theta}\Big[\frac{r^2+a^2}{\sin^2{\theta}}\nonumber\\&+&\frac{a^2\left(2Mr+2M_Dr\log(1+\frac{r_s}{r})^{\left(1+\frac{r}{r_s}\right)}\right)}{\Sigma}\Big],\\
g_{\theta\theta}&=&\Sigma,
\end{eqnarray}
\end{subequations}
where $M_{D}=\frac{4\pi \rho_sr_s^3}{1+\frac{r_s}{r}}$ is the mass profile of the DM halo and:
\begin{subequations}\label{eq.metric components2}
\begin{align}
& \Sigma=r^2+a^2\cos^2{\theta},\\
& \Delta=r^2-2Mr+a^2-2M_Dr\log(1+\frac{r_s}{r})^{(1+\frac{r}{r_s})}.
\end{align}
\end{subequations}
\begin{figure*}[ht!]\centering
\includegraphics[width=0.45\textwidth]{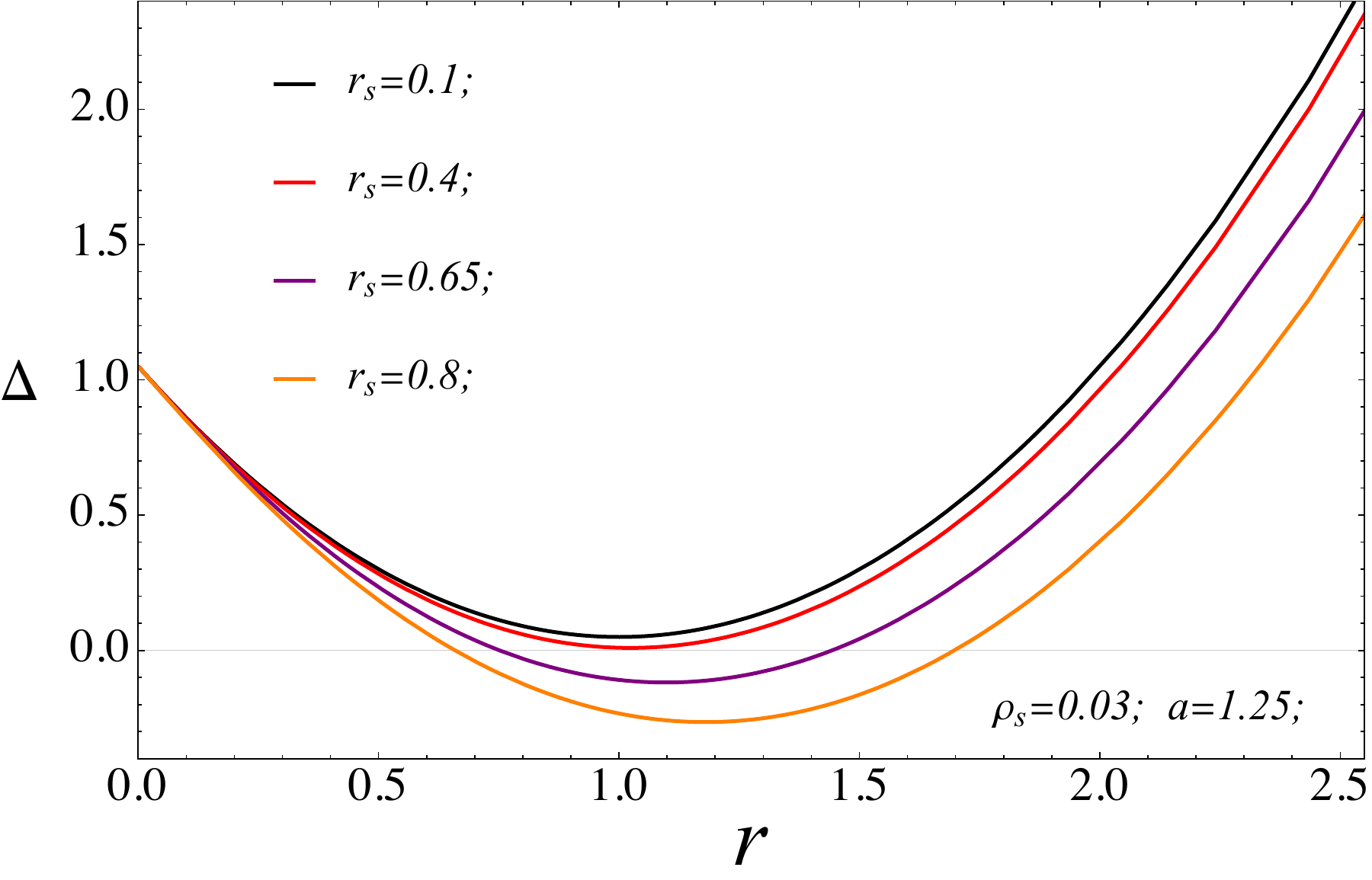}
\includegraphics[width=0.45\textwidth]{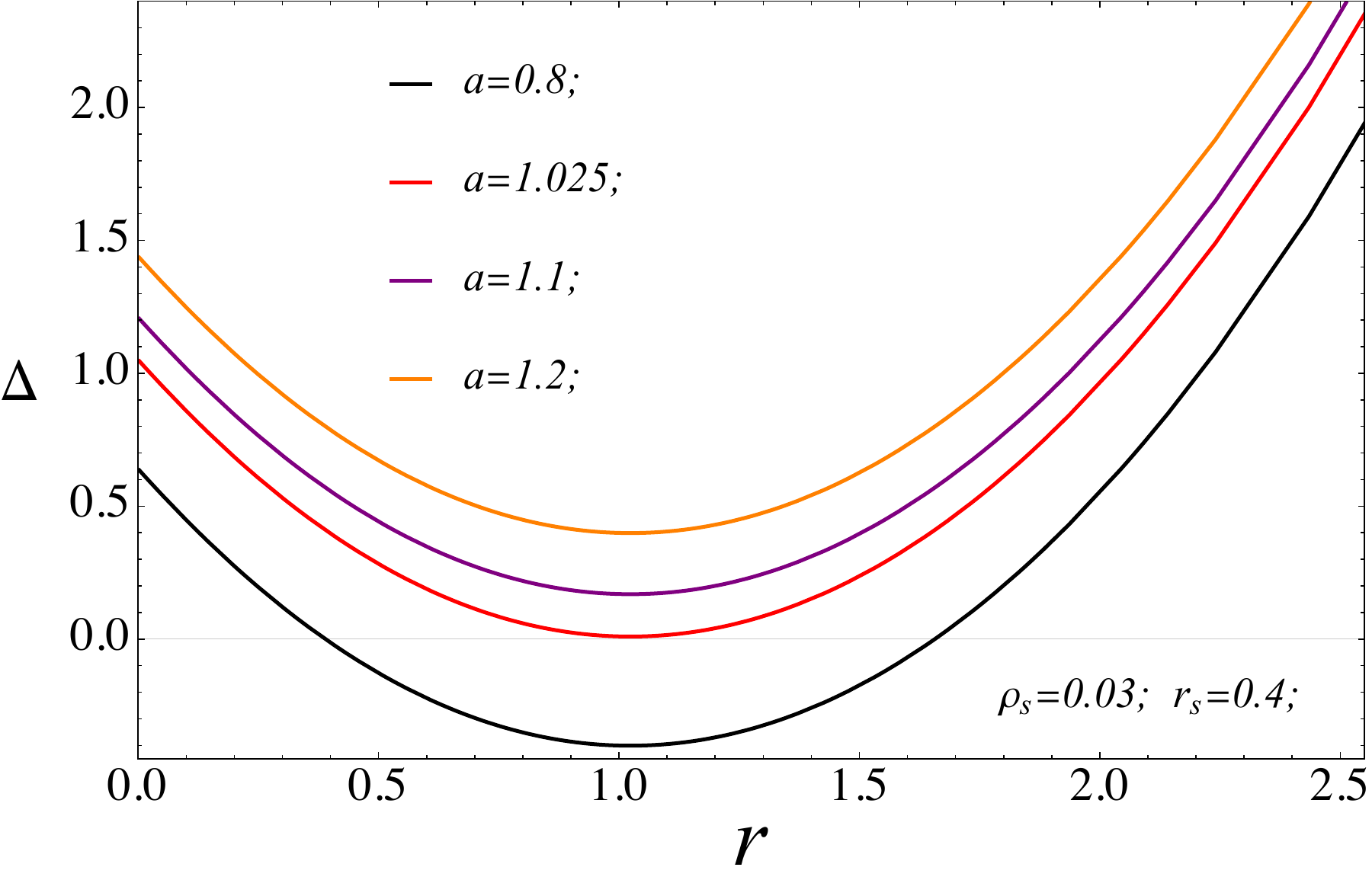}
\includegraphics[width=0.45\textwidth]{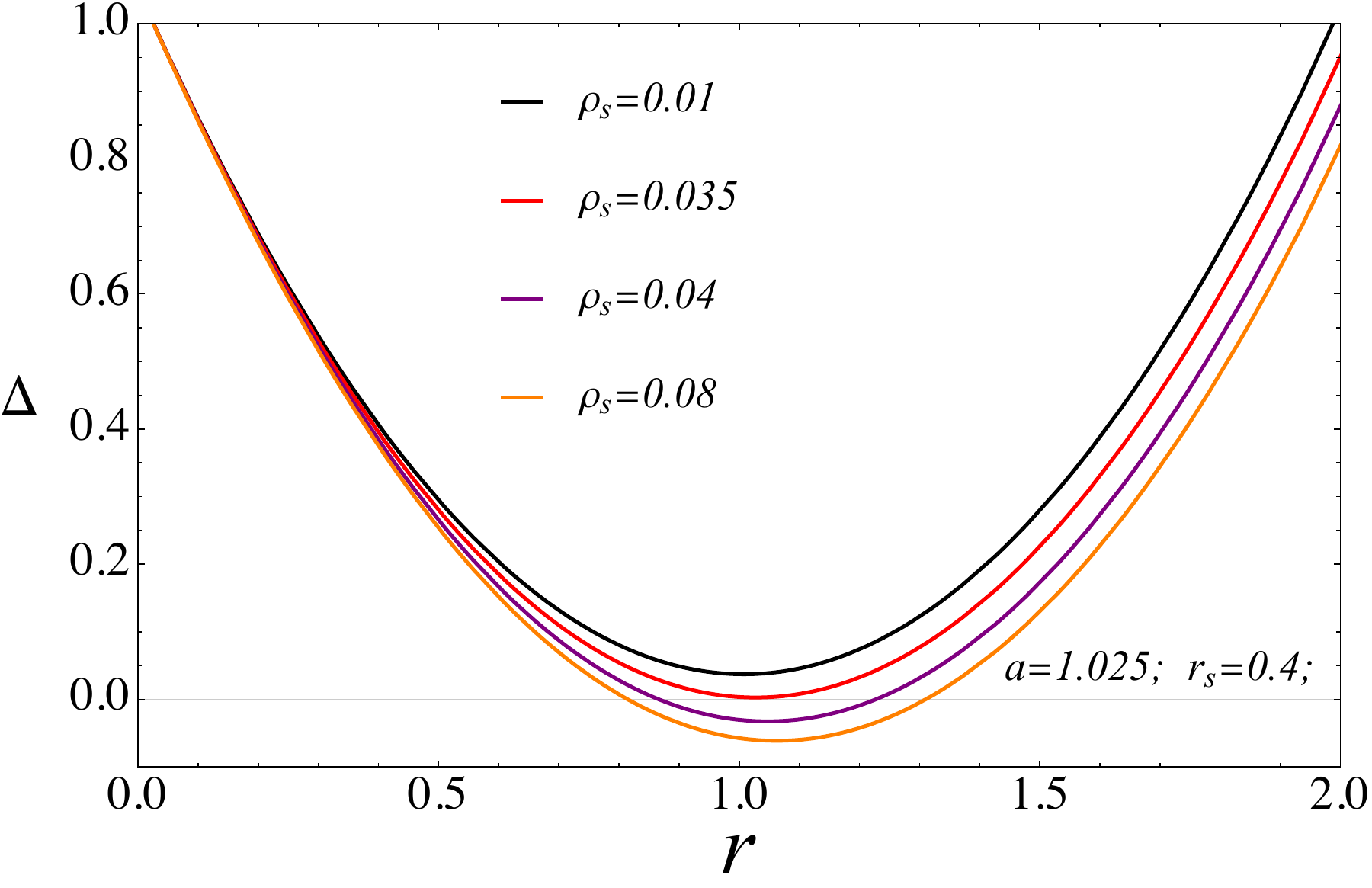}
\caption{Radial dependence $\Delta$ function for different values of the BH parameters $a$, $r_s$, $\rho_s$
\label{Fig.Delta}}
\end{figure*}
\begin{figure*}[ht!]
\includegraphics[width=0.45\textwidth]{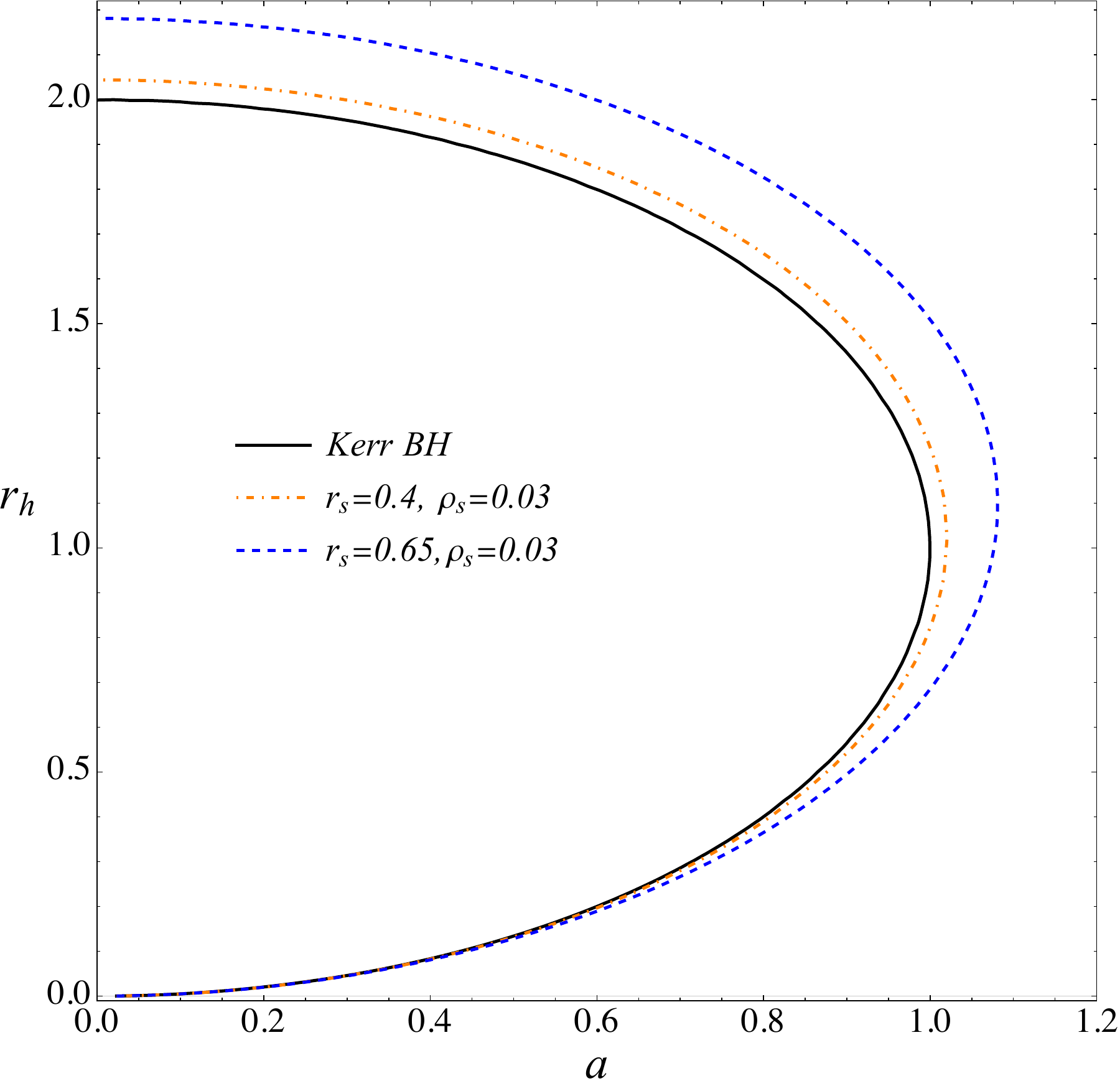}
\includegraphics[width=0.45\textwidth]{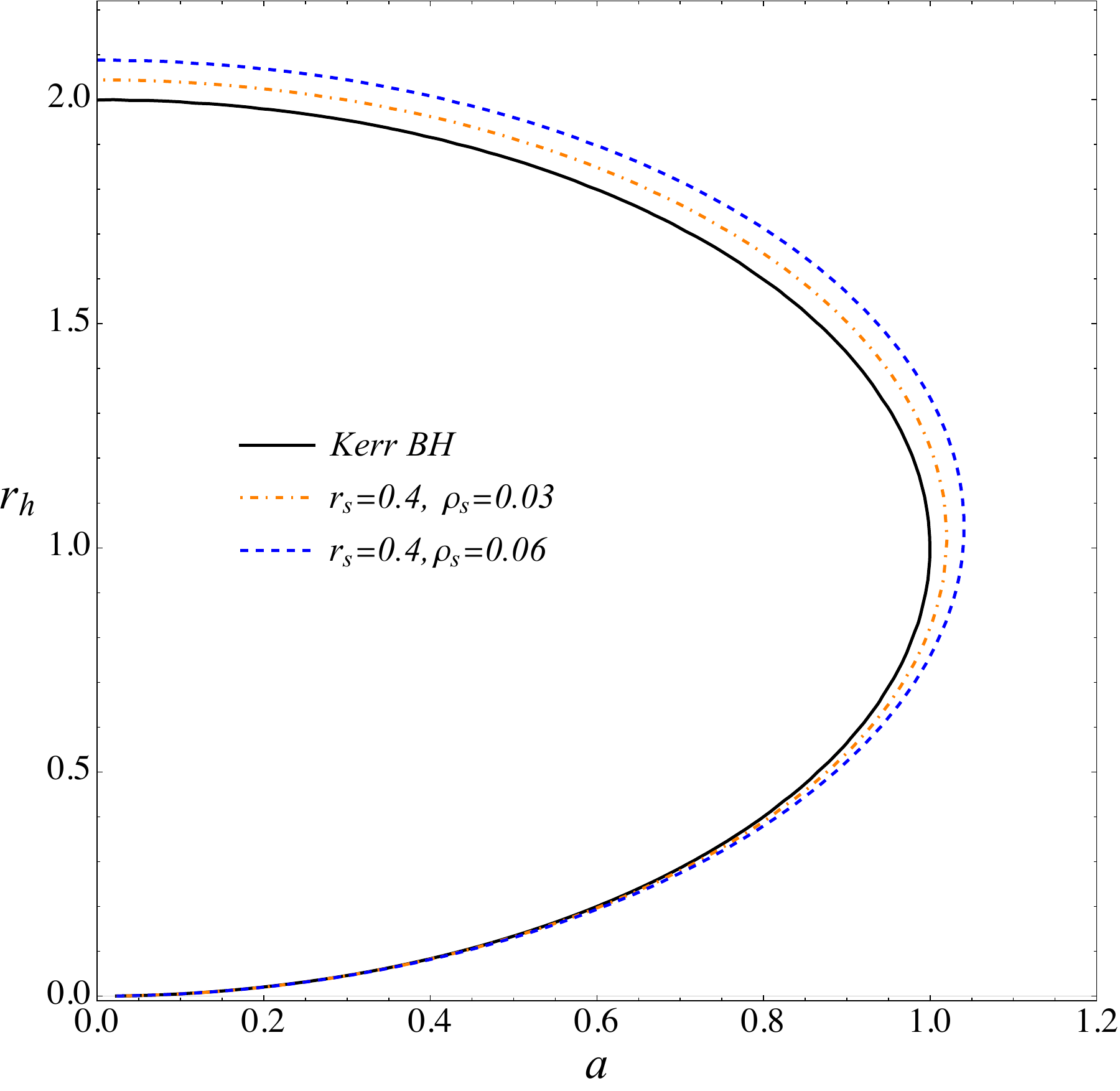}
\caption{Variation of horizon radius with increase of BH spin parameter $a$ for the selected values of parameters $\rho_s$, $r_s$. Hereafter, we shall for simplicity set $M = 1$.
\label{Fig.horizon}}
\end{figure*}
In Fig.~\ref{Fig.Delta}, we have plotted the radial dependence of the $\Delta$ function. From these graphics plotted in Fig.~\ref{Fig.Delta}, one can easily conclude, that there are two horizons for larger values of the BH parameters $r_s$ and $\rho_s$ from the top panels of Fig.~\ref{Fig.Delta}, but enlarging spin $a$ of the BH causes the disappearance of the event horizon of the BH, from the bottom panel of Fig.(\ref{Fig.Delta}).  
Then, using condition $\Delta=0$, we have plotted the dependence of the event horizon $r_h$ on the spin parameter $a$ for different values of characteristic density $\rho_s$ and characteristic scale factor $r_s$ of the DM halo in Fig. \ref{Fig.horizon}. One can easily notice that, from this graphic, increasing both BH parameters $r_s$ and $\rho_s$ causes an enlargement of the values of the radii of the event horizon. This phenomenon can be interpreted as increasing the BH parameters $r_s$ and $\rho_s$ causes enlargement of the mass profile of the DM halo, which leads to enhanced gravitational mass.

\subsection{Checking validity of the obtained rotating black hole solution}

Now our aim is to check whether our obtained rotating black hole solution with dark matter halo satisfies the Einstein field equations or not.  
Using the line element of the black hole surrounded by Dehnen-type DM halo (\ref{eq.metric components},\ref{eq.metric components2}), the components of the  Einstein tensor can be expressed as:

\allowdisplaybreaks 
\begin{strip}
\begin{subequations}\label{Eq.Einstein comp.}
    \begin{align}
&G_{tt}=-\frac{4F^2+2r\left[r^2+a^2\left(2-\cos^2{\theta}\right)\right]F_{,r}-2F\left[r^2+a^2\left(2-\cos^2{\theta}\right)+2rF_{,r}\right]-a^2\sin^2{\theta}\Sigma F_{,rr}}{\Sigma^3},\\
    &G_{rr}=-\frac{2\left(F-rF_{,r}\right)}{\Sigma\Delta},\,\,G_{\theta\theta}=\frac{2\left(F-rF_{,r}\right)+\Sigma F_{,rr}}{\Sigma},\\
    &G_{t\phi}=-\frac{a\sin^2{\theta}\left[4F\left(a^2+r^2+rF_{,r}\right)-4F^2-\left(a^2+r^2\right)\left(4rF_{,r}-\Sigma^3F_{,rr}\right)\right]}{\Sigma^3},\\
    &G_{\phi\phi}=-\frac{\sin^2{\theta}}{\Sigma^3}\{4a^2\sin^2{\theta}F^2-F\Big[2(a^2+r^2)(r^2+a^2[2-\cos^2{\theta}])\\\nonumber
&+4a^2r\sin^2{\theta}F_{,r}\Big]+(a^2+r^2)\left[2rF_{,r}(r^2+a^2(2-\cos^2{\theta}))-(a^2+r^2)\Sigma F_{,rr}\right]\},
    \end{align}
\end{subequations}
\end{strip}

where we have introduced a new variable as $F(r)=\frac{r^2\left[1-f(r)\right]}{2}$. The Einstein field equations $G_{\mu\nu}=8\pi T_{\mu\nu}$ could then be checked using an appropriate form of orthogonal bases (\cite{2014PhRvD..90f4041A,2019PhRvD.100d4012J}) such as:
\begin{subequations}\label{eq.tetrads}
    \begin{align}
        &e_t^\mu=-\frac{\left(r^2+a^2,0,0,a\right)}{\sqrt{\Sigma\Delta}}\,,\,\,e_r^\mu=-\frac{\sqrt{\Delta}\left(0,1,0,0\right)}{\sqrt{\Sigma}}\,,\\
        &e_\theta^\mu=-\frac{\left(0,0,1,0\right)}{\sqrt{\Sigma}}\,,\,\,e_\phi^\mu=\frac{\left(a\sin^2{\theta},0,0,1\right)}{\sqrt{\Sigma}\sin{\theta}}\,.
    \end{align}
\end{subequations}
Also, according to (\cite{2002PhRvD..65f4039B,2014PhRvD..90f4041A}) the components of the energy-momentum tensor $T_\mu^\nu=\text{diag}\left[-\rho,p_r,p_\theta,p_\phi\right]$ can be expressed as:
\begin{subequations}\label{eq.rho,p}
    \begin{align}
        &-\rho=p_r=\frac{\left(rF_{,r}-F\right)}{4\pi\Sigma^2}\,,\,\,p_{\theta}=p_\phi=-p_r+\frac{ F_{,rr}}{8\pi\Sigma}.
    \end{align}
\end{subequations}
Then one can easily notice that the newly obtained rotating black hole solution with a DM halo (\ref{eq.metric components}) satisfies the Einstein field equations indicating that the obtained line element is physically valid as $-\rho=\frac{1}{8\pi}e_t^\mu e_t^\nu G_{\mu\nu}$, $p_r=\frac{1}{8\pi}e_r^\mu e_r^\nu G_{\mu\nu}=\frac{1}{8\pi}g^{rr}G_{rr}$, $p_{\theta}=\frac{1}{8\pi}e_{\theta}^\mu e_{\theta}^\nu G_{\mu\nu}=\frac{1}{8\pi}g^{\theta\theta}G_{\theta\theta}$, $p_{\phi}=\frac{1}{8\pi}e_{\phi}^\mu e_{\phi}^\nu G_{\mu\nu}$.
\section{Shadow of rotating black hole in Dehnen-type DM halo}\label{Sec.III}

To find the apparent shadow of the rotating BH with a Dehnen-type DM halo, we start by studying null geodesics. The equation of motion for a photon can be expressed with the Hamilton-Jacobi equations:
\begin{eqnarray}\label{eq.HJ}
    \frac{\partial S}{\partial \lambda}=-\frac{1}{2}g^{\mu\nu}\frac{\partial S}{\partial x^\mu}\frac{\partial S}{\partial x^\nu},
\end{eqnarray}
where $\lambda$ is the affine parameter. Then, using the separability of the Hamilton-Jacobi action:
\begin{eqnarray}\label{eq.action}
    S=-Et+L\phi+S_{r}(r)+S_{\theta}(\theta),
\end{eqnarray}
where $E$ and $L$ are the energy and angular momentum of the particles, we can find the equation of motion as:
\begin{subequations}\label{eq.Equation of the motion}
\begin{align}
& \Sigma\frac{dt}{d\lambda}=a\left(L-aE\sin^2{\theta}\right)+\frac{r^2+a^2}{\Delta}\left[E(r^2+a^2)-aL\right],
\\ 
& \Sigma\frac{dr}{d\lambda}=\pm\sqrt{R} ,
\\
&  \Sigma\frac{d\theta}{d\lambda}=\pm\sqrt{\Theta},
\\
& \Sigma\frac{d\phi}{d\lambda}=\left(L\csc^2{\theta}-aE\right)+\frac{a}{\Delta}\left[E(r^2+a^2)-aL\right],
\end{align}
\end{subequations}
in which
\begin{subequations}\label{eq.part.}
\begin{align}
&R=\left[(r^2+a^2)E-aL\right]^2-\Delta\left[\mathcal{K}+(L-aE)^2\right],
\\
&\Theta=\mathcal{K}+\cos^2{\theta}\left(a^2E^2-\frac{L^2}{\sin^2{\theta}}\right) \ . 
\end{align}
\end{subequations}
Here, $\mathcal{K}$ is the separation constant.
The following effective potential can be written as 
\begin{subequations}\label{eq.part.}
\begin{align}
   & \left(\Sigma\frac{dr}{d\lambda}\right)^2+V_{eff}=0,\\
   &\frac{1}{E^2}V_{eff}=\Delta\left[\eta+(\xi-a)^2\right]-\left[(r^2+a^2)-a\xi\right]^2,
\end{align}
\end{subequations}
where we have introduced new variables $\eta=\frac{\mathcal{K}}{E^2}$ and $\xi=\frac{L}{E}$. Then we can use condition for unstable circular orbits $V_{eff}(r_{ph})=\frac{\partial V_{eff}}{\partial r}\mid_{r_{ph}}=0$, where $r_{ph}$ \cite{2024NatSR..1426932K} is the radius of the photon sphere, to find $\eta$ and $\xi$ as:
\begin{subequations}\label{eq.part.}
\begin{align}
&\xi=\frac{\Delta'\left(r^2+a^2\right)-4r\Delta}{a\Delta'},\\
&\eta=\frac{r^2\left[16\Delta(a^2-\Delta)+8r\Delta\Delta'-r^2\Delta'^2\right]}{a^2\Delta'^2}\ . 
\end{align}
\end{subequations}
Here, the prime $'$ denotes the derivative with respect to the radial coordinate $r$. 

We will introduce new celestial coordinate systems $\alpha$ and $\beta$ to investigate the dark region in the sky, the so-called black hole shadow:
\begin{subequations}\label{eq.alpha beta}
\begin{align}
&\alpha=\lim_{r\to\infty}{\left(-r^2\sin{\theta}\frac{d\phi}{dr}\mid_{\theta=\theta_0}\right)}=-\xi\csc{\theta_0},\\
&\beta=\lim_{r\to\infty}\left({r^2\frac{d\theta}{dr}\mid_{\theta=\theta_0}}\right)=\pm\sqrt{a^2\cos^2{\theta}+\eta-\xi^2\cot^2{\theta}},
\end{align}
\end{subequations}
where $\theta_0$ is the angle of inclination of the observer. For the BH shadow in the equatorial plane $(\theta_0=\frac{\pi}{2})$ Eqs.~(\ref{eq.alpha beta}) can be rewritten in the form:
\begin{subequations}\label{eq.alpha beta 2}
\begin{align}
&\alpha=-\xi,\\
&\beta=\pm\sqrt{\eta}.
\end{align}
\end{subequations}

Now, to obtain the silhouette of the BH shadow, one may plot the line described by $\xi$ vs. $\eta$ in celestial coordinates \cite{Tsukamoto:2014tja}. 
\begin{figure*}[ht!]\centering
\includegraphics[width=0.45\textwidth]{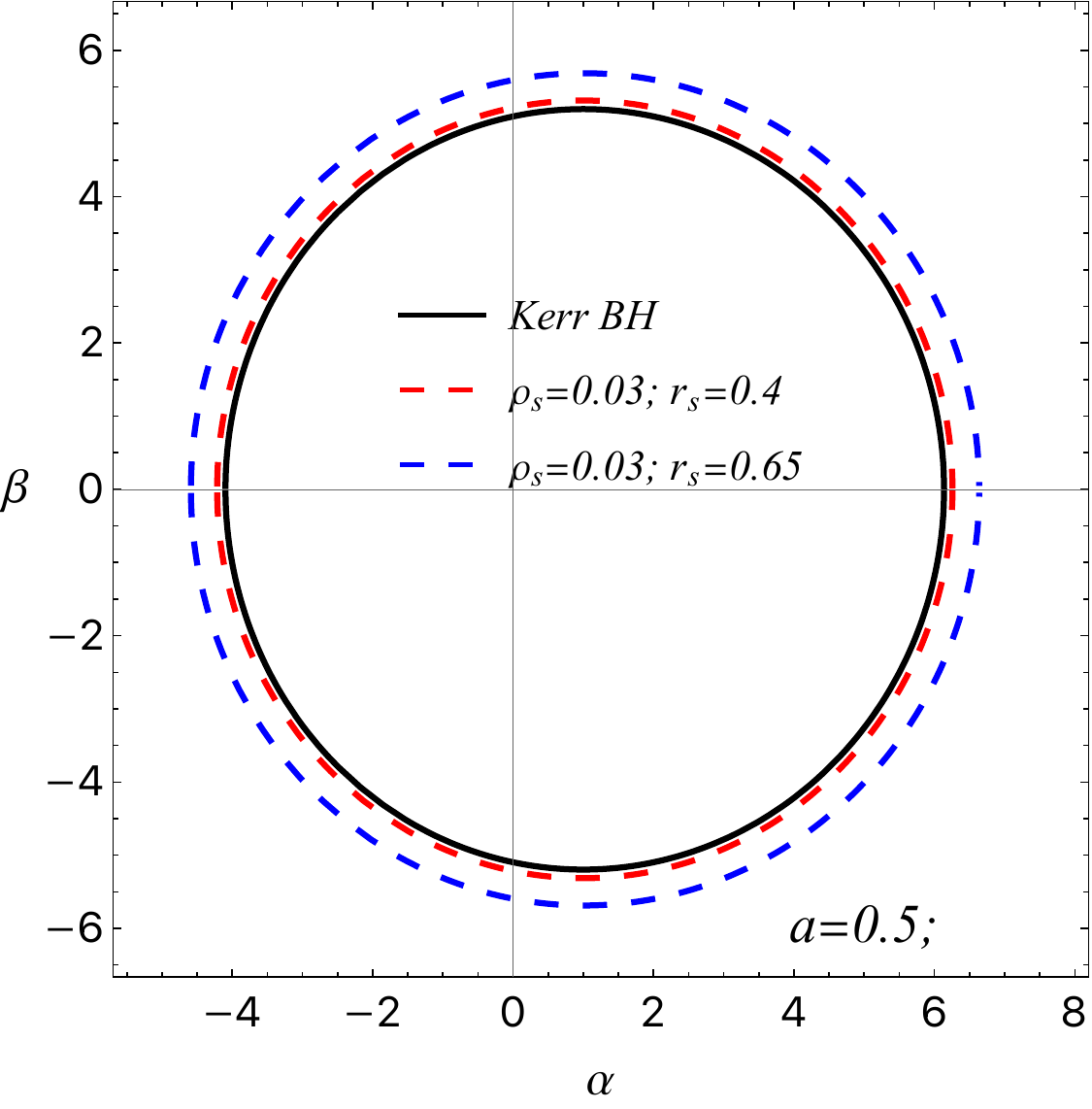}
\includegraphics[width=0.45\textwidth]{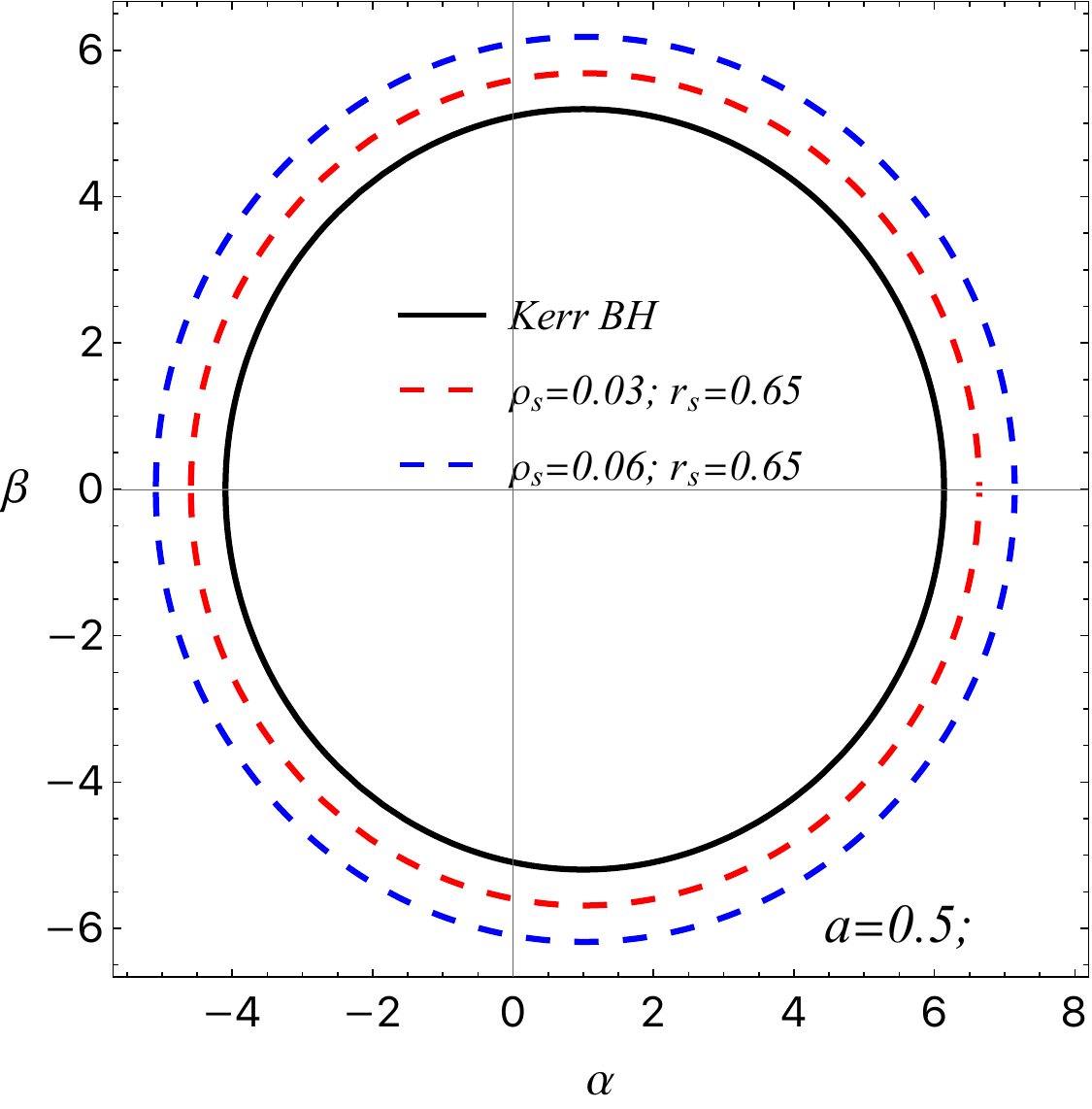}
\includegraphics[width=0.45\textwidth]{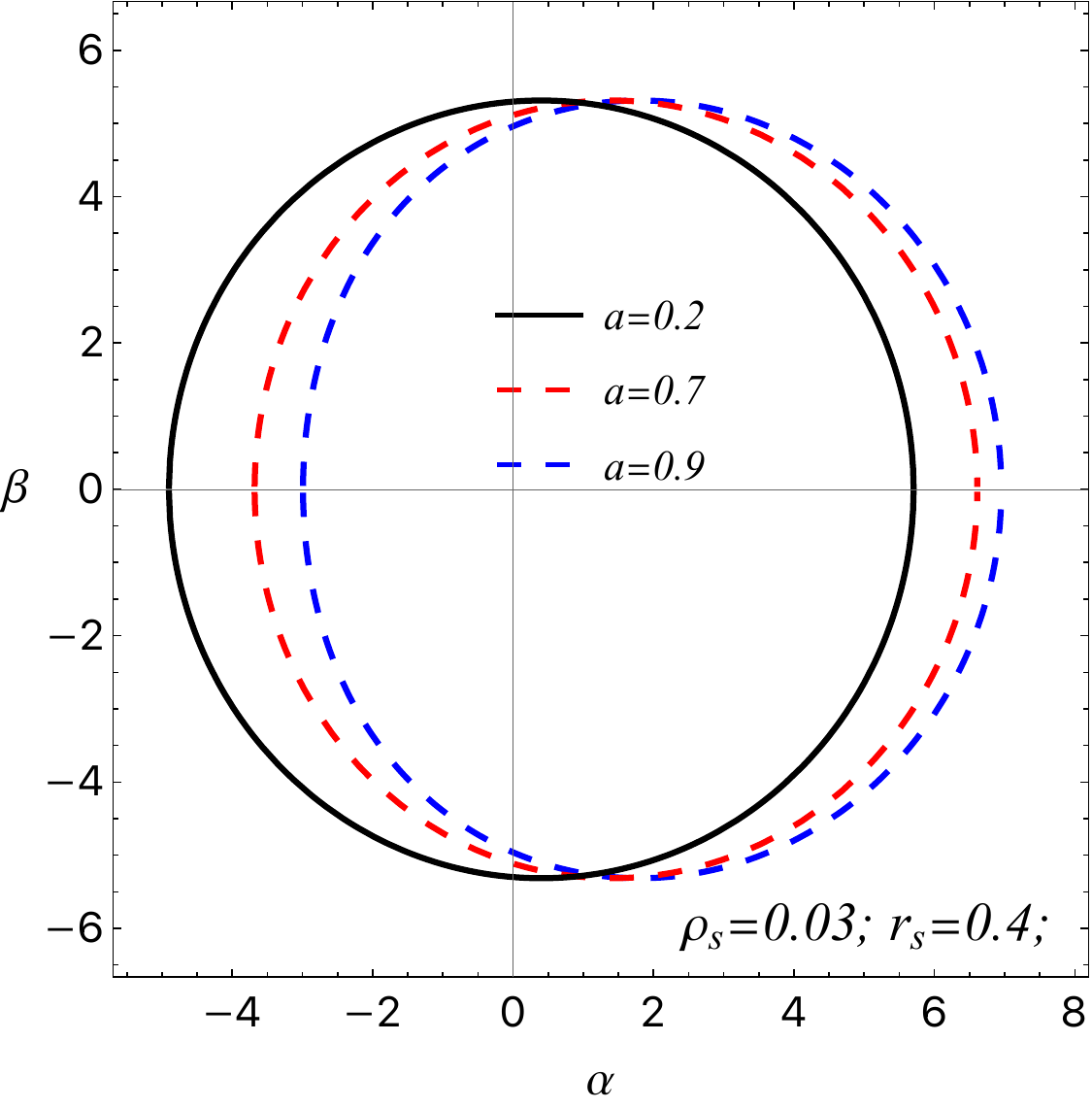}
\includegraphics[width=0.45\textwidth]{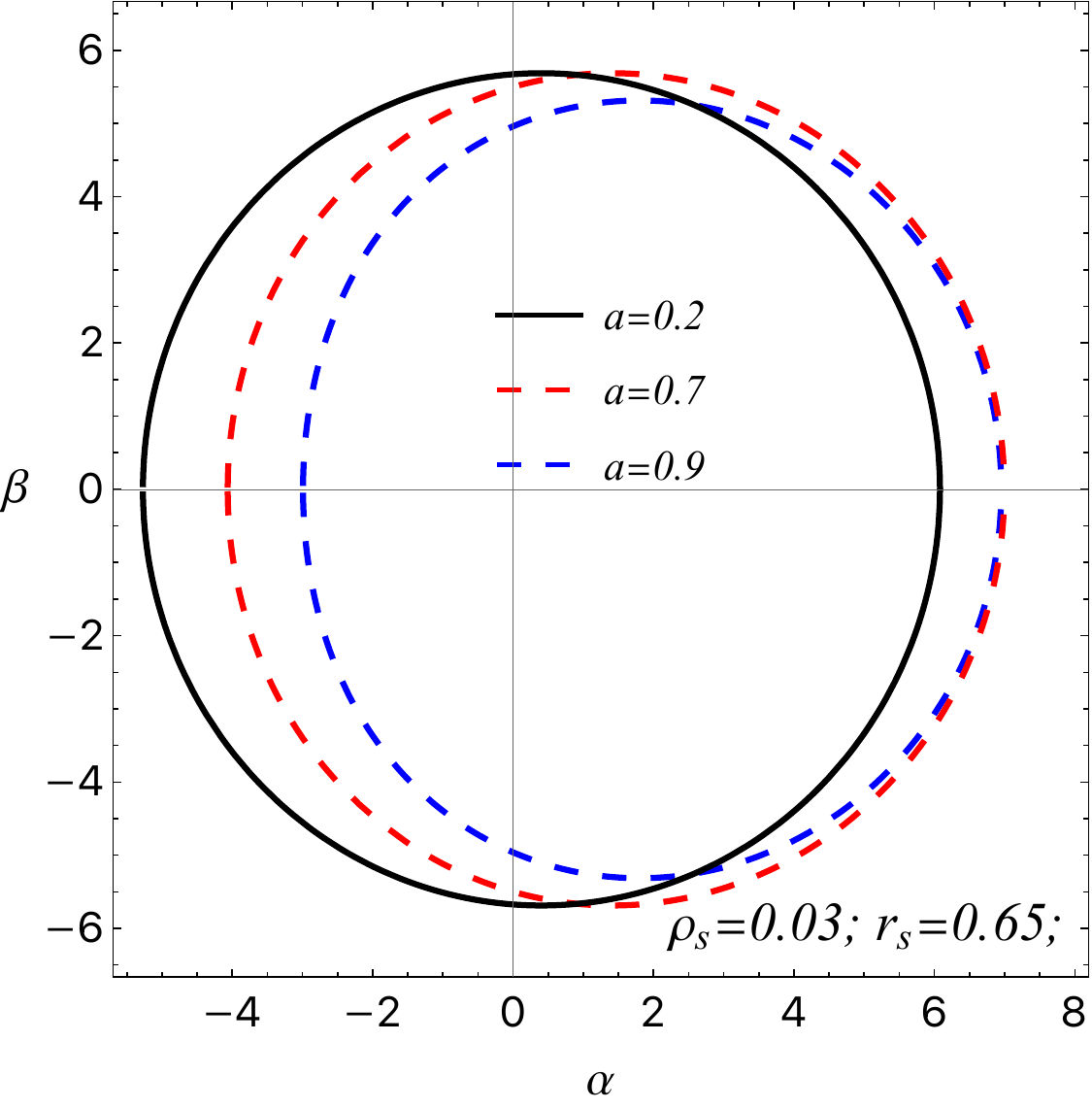}
\caption{BH shadow for various  of the BH parameters $r_s$, $\rho_s$, $a$.
\label{Fig.shadow}}
\end{figure*}
Then, in Fig.~\ref{Fig.shadow}, we have shown the shadow of the rotating BH surrounded by a Dehnen-type dark matter halo. Again, from these graphs shown in~\ref{Fig.shadow}, one can easily conclude that increasing the values of the BH parameters $r_s$ and $\rho_s$ leads to an increase in the shadow of the BH. In addition, the well-known effect of the spin $a$ of the BH on the shadow, distortion of the shadow of the BH,  is seen from the bottom panels of the graphics plotted in~\ref{Fig.shadow}.  

\begin{figure*}[ht!]\centering
\includegraphics[width=0.45\textwidth]{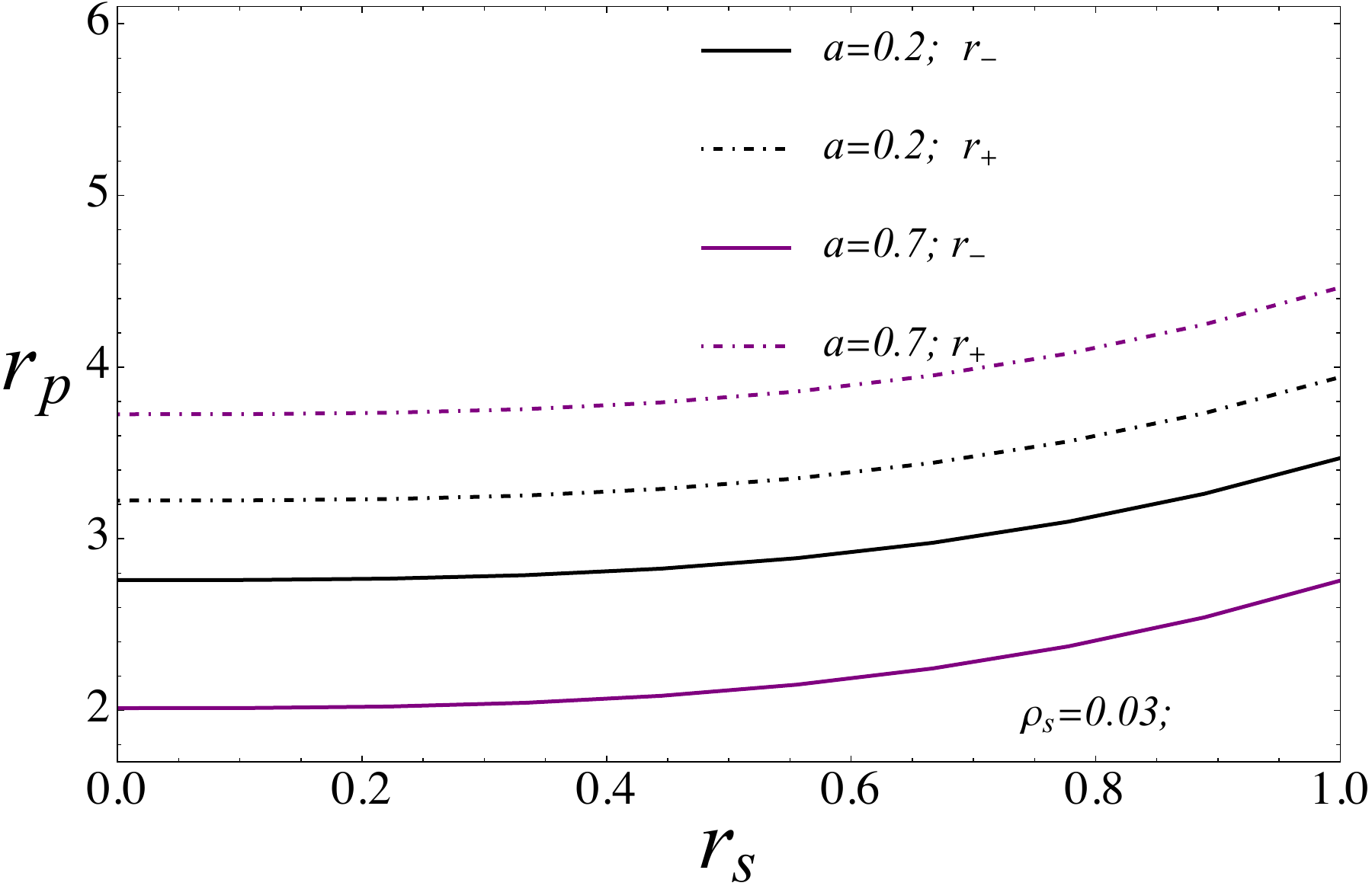}
\includegraphics[width=0.45\textwidth]{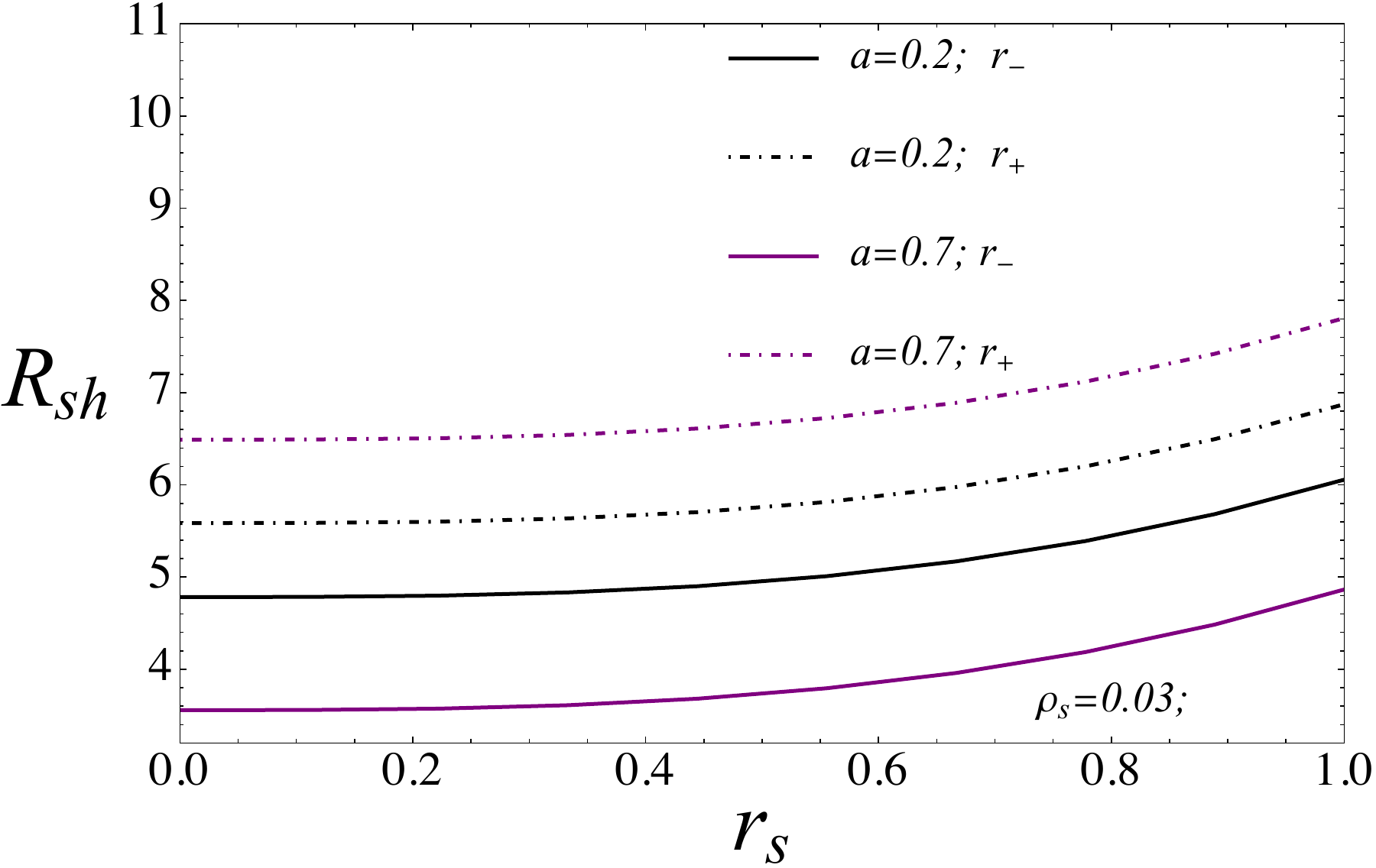}
\includegraphics[width=0.45\textwidth]{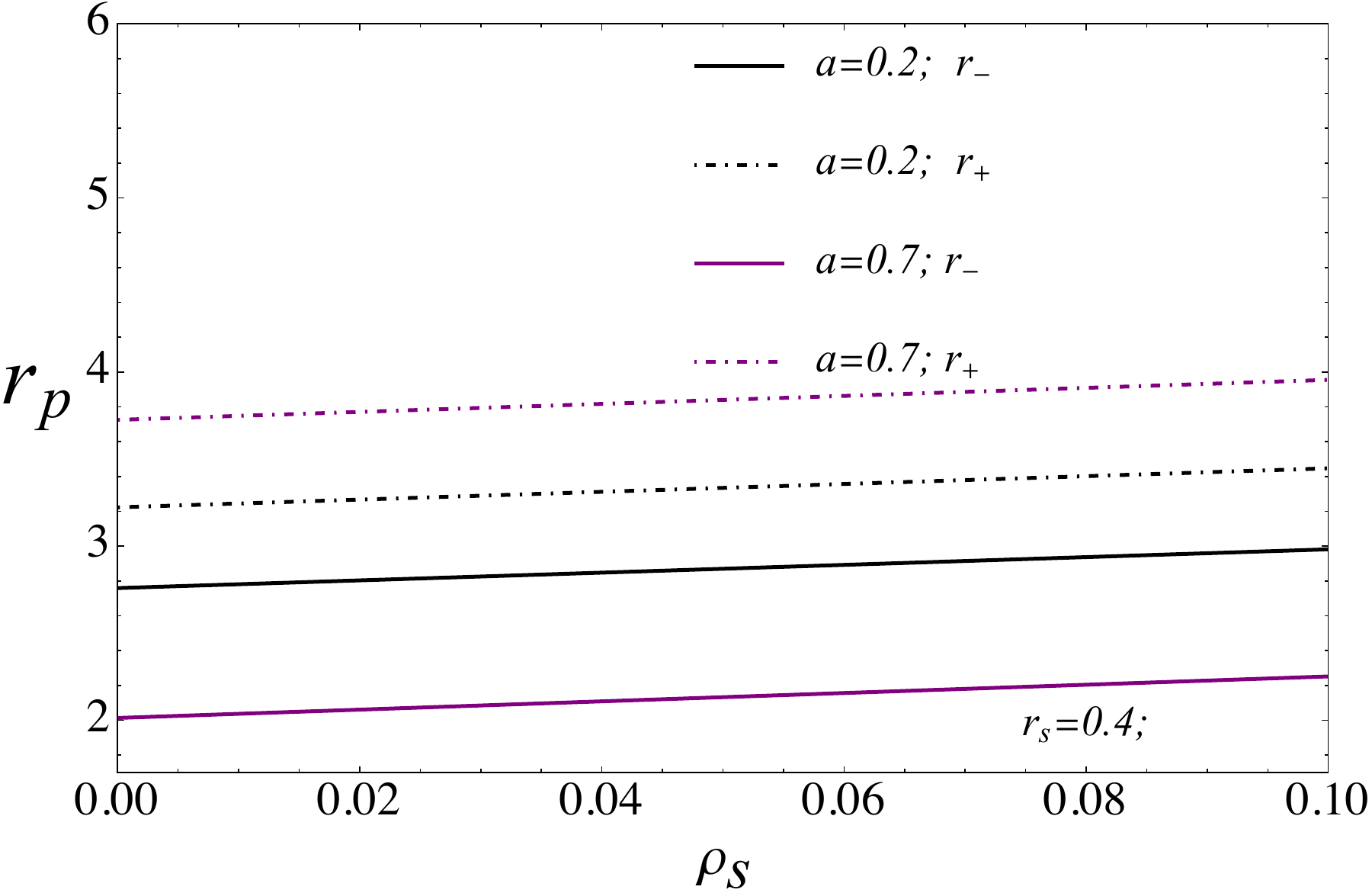}
\includegraphics[width=0.45\textwidth]{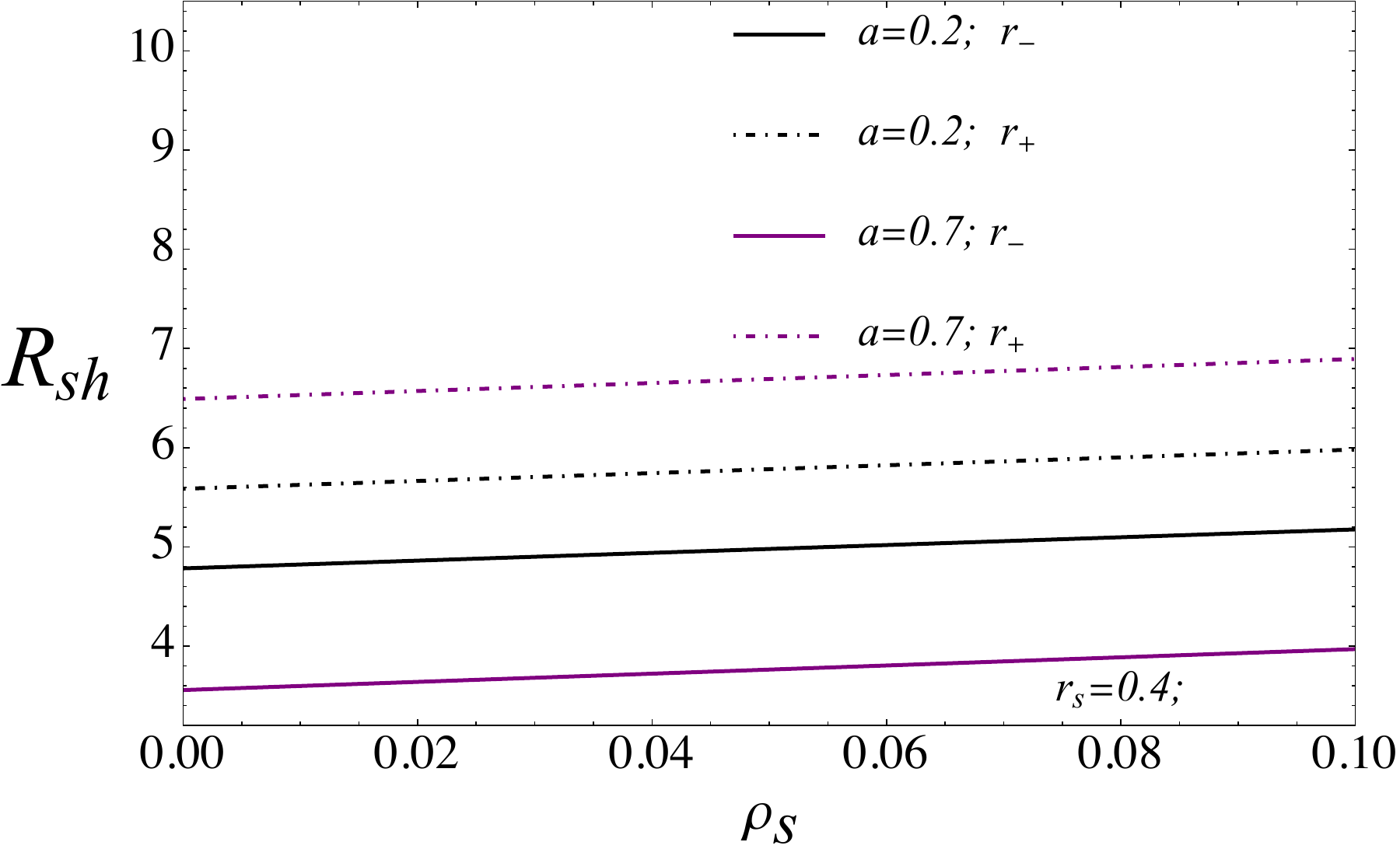}
\caption{Dependence of the shadow size $R_{sh}$ and photon sphere $r_p$ with respect to different spacetime parameters, here $r_{\pm}$ refers to the prograde and retrograde radii of stable circular orbits. 
\label{Fig.shadow radius}}
\end{figure*}

The expression for the shadow radius can be expressed as:
\begin{eqnarray}\label{eq.shadow radius}
    R_{sh}^2=\alpha^2+\beta^2,
\end{eqnarray}
Then, in Fig.~\ref{Fig.shadow radius} we have plotted the photon sphere and shadow radius for different spacetime parameters. 

One can notice from Fig.~\ref{Fig.shadow radius} enlarging the values of both space-time parameters $r_s$ and $\rho_s$ causes increasing the radii of photon sphere $r_p$ and the size of the shadow $R_{sh}$, which is coincide with the interpretation of the Fig.~\ref{Fig.shadow} we have done in above discussion. 

\section{Deflection angle of light by a rotating black hole in Dehnen-type DM halo.}\label{Sec.4}

\begin{figure*}[ht!]\centering
\includegraphics[width=0.5\textwidth]{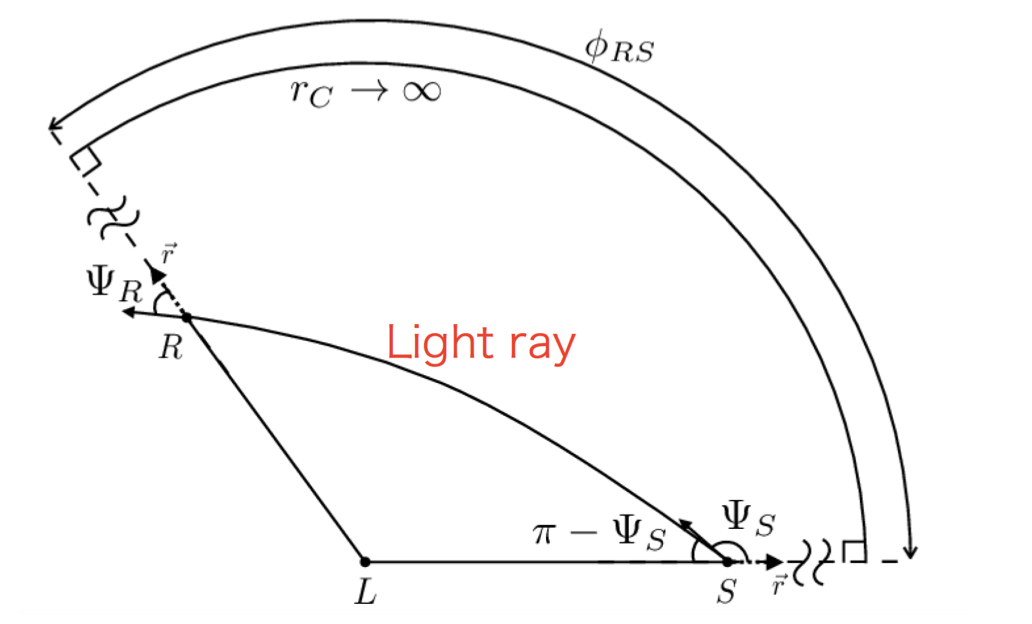}
\caption{The geometry of a quadrilateral embedded in curved space is presented.\cite{Ovgun:2018fte} 
\label{Fig.light bend}}
\end{figure*}

In this section, we will study the deflection angle for a rotating black hole in Dehnen-type DM halo via OIA method (\cite{Ono:2017pie,Ovgun:2018fte}). To achieve this, we begin by applying the null condition 
$ds^2=0$ and derive the orbital equation in the equatorial plane, where $\theta=\frac{\pi}{2}$. We also utilize a bounded, two-dimensional orientable surface, as depicted in Fig. \ref{Fig.light bend}. Following, the geodesic constants, energy $E=-k_{(t)}^{\mu}p_\mu$ and angular momentum $L=k_{(\phi)}^{\mu}p_{\mu}$in which:
\begin{subequations}
    \begin{align}
        & k_t^\mu=(1,0,0,0)\,\,\,\,\text{timelike Killing vector}\,,\nonumber\\
        &k_\phi^\mu=(0,0,0,1)\,\,\,\,\text{spacelike Killing vector}\,,\nonumber
    \end{align}
\end{subequations}
for the line element (\ref{eq.metric components}) can be expressed as:
\begin{subequations}
    \begin{align}
        & E=-g_{tt}\Dot{t}-g_{t\phi}\Dot{\phi}\,,\\
        &L=g_{t\phi}\Dot{t}+g_{\phi\phi}\Dot{\phi}\,,
    \end{align}
\end{subequations}
here dot means derivative with respect to affine parameter.

Subsequently, impact parameter $b$ can be defined as:
\begin{eqnarray}\label{eq.b}
    b=\frac{L}{E}=\frac{g_{t\phi}+g_{\phi\phi}\frac{d\phi}{dt}}{-g_{tt}-g_{t\phi}\frac{d\phi}{dt}}\,,
\end{eqnarray}
which paves a way to find $\frac{dr}{d\phi}$ using condition $ds^2=0$ for the line element (\ref{eq.metric components}) as:
\begin{eqnarray}\label{eq.dr}
 (\frac{dr}{d\phi})^2=\frac{(g_{t\phi}^2-g_{tt}g_{\phi\phi})(b^2g_{tt}+2bg_{t\phi}+g_{\phi\phi})}{g_{rr}(g_{tt}b+g_{t\phi})^2}\,,   
\end{eqnarray}
or after introducing new variable as $u=\frac{1}{r}$ Eq. (\ref{eq.dr}) can be rewritten as:
\begin{eqnarray}\label{eq.u}
 (\frac{du}{d\phi})^2=\frac{u^4(g_{t\phi}^2-g_{tt}g_{\phi\phi})(b^2g_{tt}+2bg_{t\phi}+g_{\phi\phi})}{g_{rr}(g_{tt}b+g_{t\phi})^2}\,. 
\end{eqnarray}

Again, using the null condition $ds^2=0$ for metric (\ref{eq.metric components}) we are able to find $dt$ as:
\begin{subequations}\label{eq.dt}
\begin{align}
       &dt=\sqrt{\gamma_{ij}dx^idx^j}+\beta_idx^i\,,\\
       &dl^2=\gamma_{ij}dx^idx^j=\frac{g_{t\phi}^2-g_{\phi\phi}g_{tt}}{g_{tt}^2}d\phi^2-\frac{g_{rr}}{g_{tt}}dr^2\,,\\
       &\beta_idx^i=-\frac{g_{t\phi}}{g_{tt}}d\phi\,,
    \end{align}
\end{subequations}
in which $\gamma_{ij}\neq g_{ij}\,\,(i,j=1,2,3)$  is the spatial metric and $l$ is the arc-length. 

Then, on the equatorial plane the photon’s path is described by a unit tangent vector $e^i$, given by(\cite{Ovgun:2018fte},\cite{Ono:2017pie}):
\begin{eqnarray}\label{eq.unit vector}
    e^i=\frac{1}{\chi}\left(\frac{dr}{d\phi},0,1\right)\,,
\end{eqnarray}
in which $\chi$ can be obtained using condition $\gamma e^ie^j=1$ as
\begin{eqnarray}
    \frac{1}{\chi}=\frac{g_{tt}(g_{tt}b+g_{t\phi})}{g_{t\phi}^2-g_{tt}g_{\phi\phi}}\,.
\end{eqnarray}

In the case where the unit radial vector is directed outgoing, we find:
\begin{eqnarray}\label{eq.R}
    R^i=\left(\frac{1}{\sqrt{\gamma_{rr}},},0\,,0\,\right)\,,
\end{eqnarray}
which enables to find angle computed from the outgoing radial direction as:
\begin{eqnarray}\label{eq.sin}
    \sin{\Psi}=-\frac{g_{t\phi}+bg_{tt}}{\sqrt{g_{t\phi}^2-g_{tt}g_{\phi\phi}}}\,,
\end{eqnarray}
where we have used expression $\cos{\Psi}=\gamma_{ij}e^iR^j$. Eq.(\ref{eq.sin}) can be expressed for metric (\ref{eq.metric components2}) in the form:
\begin{eqnarray}\label{eq.sin2}
    \sin{\Psi}=\frac{b}{r}\times\frac{1-\frac{2\Tilde{M}}{r}+\frac{2a\Tilde{M}}{br}}{\sqrt{1-\frac{2\Tilde{M}}{r}+\frac{a^2}{r^2}}}\,,
\end{eqnarray}
here $\Tilde{M}=\left[M+M_D\log{(1+\frac{r_s}{r})^{1+\frac{r}{r_s}}}\right]$, eq.(\ref{eq.sin2}) converts to the exactly Kerr form (\cite{Ono:2017pie}) when $\rho_s\to0$. Subsequently, eq.(\ref{eq.sin2}) can be aproximated as:
\begin{eqnarray}\label{eq.sin3}
    \sin{\Psi}=\frac{b}{r}\left(1-\frac{\Tilde{M}}{r}+\frac{2a\Tilde{M}}{br}\right)+\mathcal{O}(\frac{\Tilde{M}^2}{r^2},\frac{a^2}{r^2})\,.
\end{eqnarray}
Then we are able to obtain relation between the angles at the receiver $\Psi_R$ and  the source $\Psi_S$ in the form:
\begin{eqnarray}\label{eq.angle relation}
    \Psi_R-\Psi_S&=&\arcsin{(bu_R)}+\arcsin{(bu_S)}-\pi-\\\nonumber
    &-&\frac{\Tilde{M}bu_R^2}{\sqrt{1-b^2u_R^2}}-\frac{\Tilde{M}bu_S^2}{\sqrt{1-b^2u_S^2}}+\\\nonumber
    &+&\frac{2\Tilde{M}u_R^2}{\sqrt{1-b^2u_R^2}}+\frac{2\Tilde{M}u_S^2}{\sqrt{1-b^2u_S^2}}\,.
\end{eqnarray}

The deflection angle can now be formulated in terms of the newly defined angles at the receiver $\Psi_R$, the source $\Psi_S$, and the coordinate angle $\phi_{RS}$ as:
\begin{eqnarray}\label{eq.angle}
    \Hat{\alpha}=\Psi_R-\Psi_S+\phi_{RS}\,,
\end{eqnarray}
where $\phi_{RS}$ can be calculated employing Eq.(\ref{eq.dr}) as:
\begin{eqnarray}\label{eq.phi}
  \phi_{RS}&=&\pi-\arcsin{(bu_R)}-\arcsin{(bu_S)}+\\\nonumber
  &+&\frac{2 \Tilde{M}^*}{b}\left[\frac{2-b^2u_R^2}{2\sqrt{1-b^2u_R^2}}+\frac{2-b^2u_S^2}{2\sqrt{1-b^2u_S^2}}\right]\\\nonumber
  &-&\frac{2a\Tilde{M}^*}{b^2}\left[\frac{1}{\sqrt{1-b^2u^2_S}}+\frac{1}{\sqrt{1-b^2u^2_S}}\right]\,,
\end{eqnarray}
here we used weak field and slow rotation approximations (see, (\cite{Ono:2017pie}), \cite{Ishihara:2016vdc} for details).

Finally, taking far limit $r_S\to0,\,\,\,r_R\to0$ and substituting Eqs.(\ref{eq.angle relation}) and (\ref{eq.phi}) into Eq.(\ref{eq.angle}) we will have expression for deflection angle:
\begin{eqnarray}\label{eq.angle2}
    \hat{\alpha}=\frac{4\Tilde{M}^*}{b}\pm\frac{4a\Tilde{M}^*}{b^2}\,,
\end{eqnarray}
where $+,\,\,\,-$ signs here correspond to the  retrograde and prograde photon orbits and $\Tilde{M}^*\approx M+4\pi\rho_sr_s^3$.
\begin{figure*}[ht!]\centering
\includegraphics[width=0.45\textwidth]{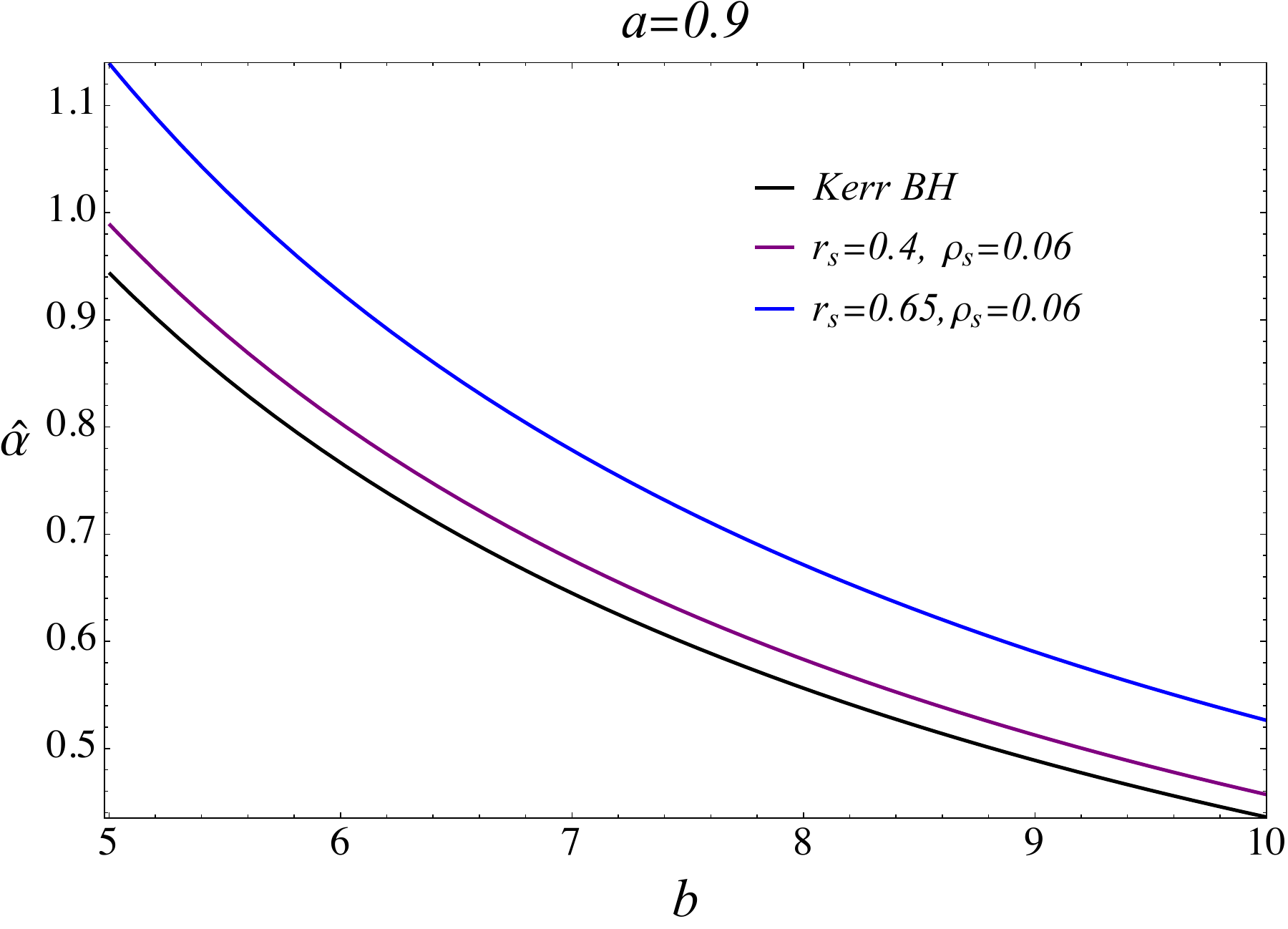}
\includegraphics[width=0.45\textwidth]{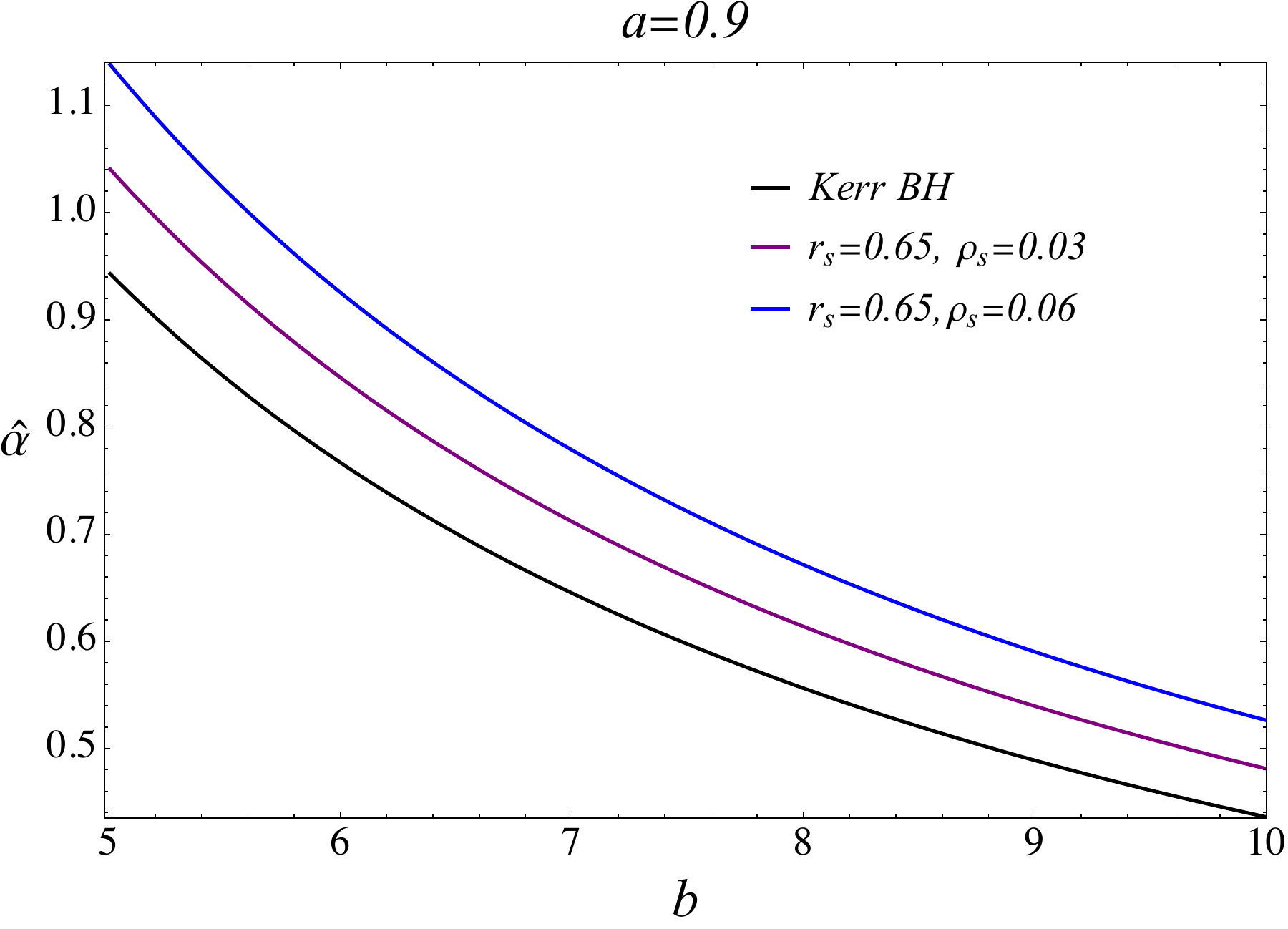}
\caption{Dependence of the deflection angle $\hat{\alpha}$ on the impact parameter $b$ for various cases. 
\label{Fig.deflection}}
\end{figure*}
Then we have shown how deflection angle $\hat{\alpha}$ depend on the impact parameter $b$ in Fig.(\ref{Fig.deflection}). One can see from graphics in Fig.(\ref{Fig.deflection}) that presence of the DM halo increases the value of the deflection angle $\hat{\alpha}$. 

\section{Constraints from EHT observational data }
\label{sec.EHT}

Now we will obtain constraints from the observational data of the EHT on the image characteristics of SMBHs. To do it, we ought to find a closed area of the  shadow silhouette as \cite{2023EPJC...83..855Z} (\cite{2015MNRAS.454.2423A}):
\begin{eqnarray}\label{eq.area}
    A=2\int_{r_-}^{r_+}\left[\beta(r)\frac{d\alpha(r)}{dr}\right]dr,
\end{eqnarray}
which enables us to find the radius of the areal shadow as $R_a=\sqrt{\frac{A}{\pi}}$. 

Now we can determine the constraints for spacetime  parameters of the supermassive black holes Sgr A$*$ and M87$*$, assuming them as rotating black holes with a dark matter halo and using the angular diameter of their shadow image measured by EHT observation at a distance $d$ from the black hole, which can be expressed as:
\begin{eqnarray}\label{eq.angular diameter}
    \Theta_d=2\frac{R_a}{d}.
\end{eqnarray}
\begin{figure*}[ht!]\centering
\includegraphics[width=0.45\textwidth]{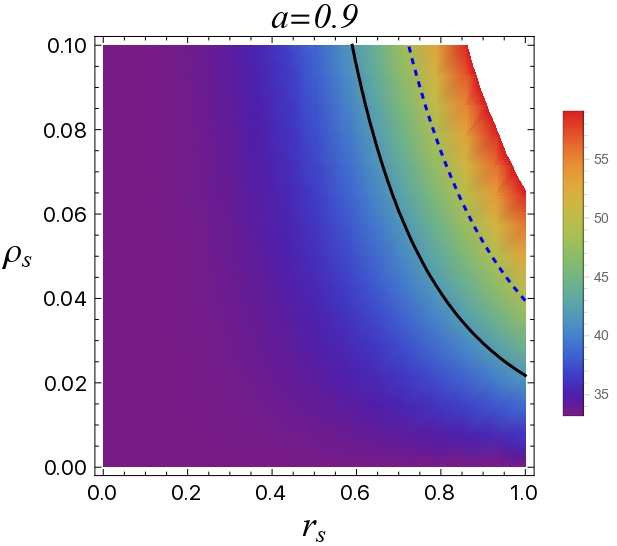}
\includegraphics[width=0.45\textwidth]{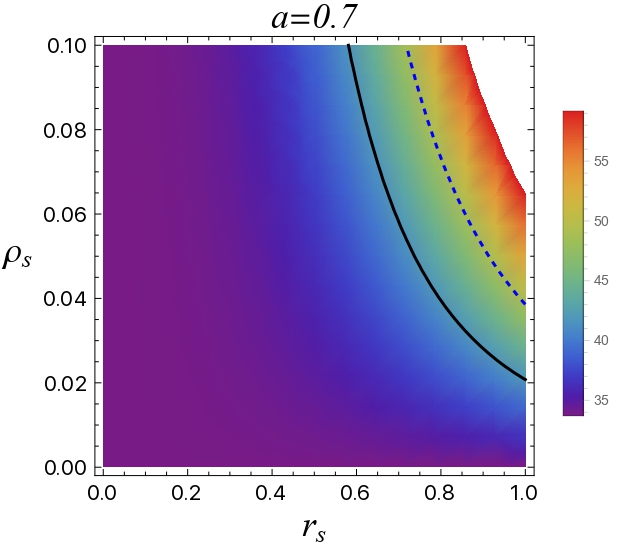}
\includegraphics[width=0.45\textwidth]{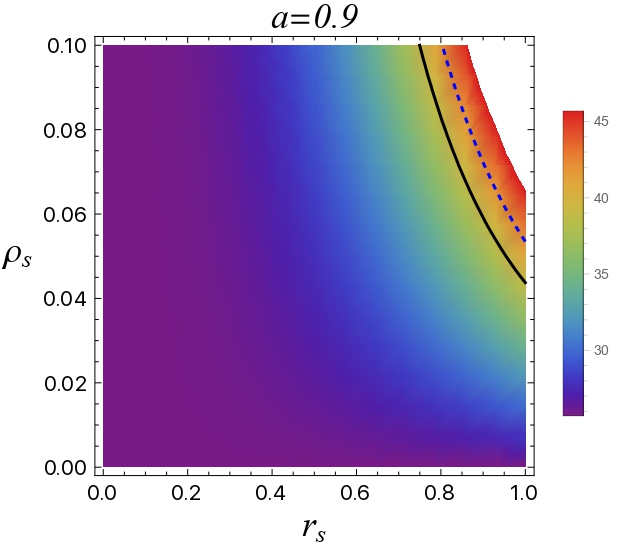}
\includegraphics[width=0.45\textwidth]{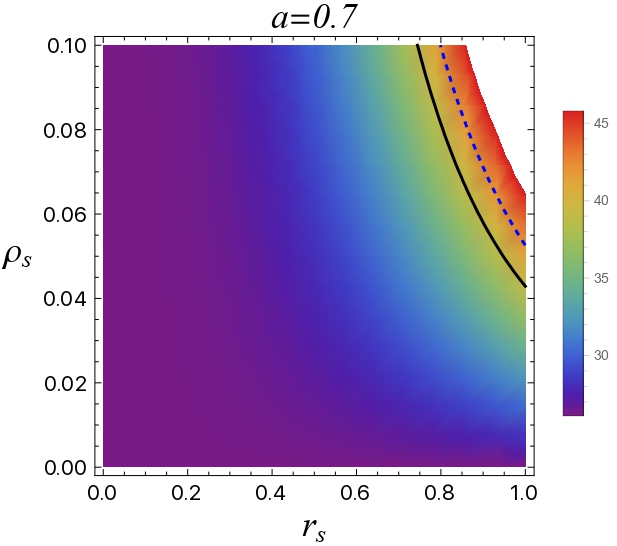}
\caption{Angular diameter observable $\Theta_d$ for the black hole shadows as a function of parameters $r_s$ and $\rho_s$. The black curves correspond to
$\Theta_d = 41.7\mu as$ (for top panels) and $\Theta_d = 39\mu as$ (for bottom panels)  within $1\sigma$ region of the measured angular diameter, $\Theta_d = 48.7 \pm 7\mu as$, of the $Sgr A*$ black hole (top panels) and $\Theta_d = 42 \pm 3\mu as$ of the $M87*$ black hole reported by the EHT. The blue dashed curve corresponds to $48.7\mu as$ and $42\mu as$, respectively.  
\label{Fig.constrain}}
\end{figure*}

The observational angular diameters of the shadow images of the SMBHs M87$*$ and Sgr A$*$ are $\Theta_d=42\pm3 \mu as$ and $\Theta_d=48.7\pm7\mu as$, respectively (\cite{2019ApJ...875L...1E},\cite{2022ApJ...930L..12E}). 
We would like to point out that, for simplicity,  we do not take into account the possible uncertainties in the mass and distance measurements of the black holes in the numerical estimations.

Then in the top panel of Fig.~\ref{Fig.constrain}, we present the density plots of the characteristic scale $r_s$ and characteristic density $\rho_s$ of the black hole that are fitted with the measured values of the angular diameter of Sgr A$*$, where we assume the mass of SMBH SgrA$*$ as $M\approx 4\times 10^{6}M_{\odot}$ and the distance from Earth to SMBH Sgr A$*$ as $d\approx 8kpc$ (\cite{2022ApJ...930L..12E}), for the different fixed values of the BH spin $a$. To be more informative, we show the lower bound $\Theta_d=41.7\mu as$ (black curves on top panels of Fig.~\ref{Fig.constrain}) and the mean value $\Theta_d=48.7\mu as$ (blue dashed lines on top panels of Fig.~\ref{Fig.constrain} of the angular diameter of the supermassive black hole Sgr A$*$. It is clear from Fig.~\ref{Fig.constrain} that the values of the BH parameters $r_s$ and $\rho_s$ should be relatively high to coincide with the EHT observational data for Sgr A $*$.

In addition, we performed the same analysis for the measured value of the angular diameter of SMBH M87$*$ (\cite{2019ApJ...875L...1E}) in the bottom panel of Fig.~\ref{Fig.constrain}. Again, we present the lower bound $\Theta_d=39\mu as$ (solid black lines in the bottom graphics of Fig.~\ref{Fig.constrain}), and mean value $\Theta_d=42$ (blue dashed lines in the bottom panels of Fig.~\ref{Fig.constrain}) of the angular diameter of the supermassive black hole M87$*$. Similarly, the characteristic scale $r_s$ and the characteristic density $\rho_s$ of the rotating BH in the dark matter halo have to be relatively high to coincide with the EHT observational data for M87$*$.

\subsection{Parameter estimation for a rotating black hole in Dehnen-type DM halo through MCMC analysis}

In this section we constrain DM parameter $\rho_s$, $r_s$ and BH mass $M$, BH spin $a$ employing MCMC analysis and using  observed data for $Sgr A^*$ from EHT collaboration (\cite{2019ApJ...875L...1E}), observed data for $Sgr A^*$ from Gravity collaboration (\cite{GRAVITY:2021xju}), observed data for $M 87^*$ from EHT collaboration (\cite{2022ApJ...930L..12E}).

We start with posterior distribution:
\begin{eqnarray}\label{eq.posterior}
    P\left(\Theta|\mathcal{D},\mathcal{M}\right)=\frac{\mathcal{L}\left(\Theta|\mathcal{D},\mathcal{M}\right)\pi\left(\Theta|\mathcal{M}\right)}{P\left(\mathcal{D},\mathcal{M}\right)}\,,
\end{eqnarray}
where the likelihood function $L\left(\Theta|\mathcal{D},\mathcal{M}\right)$ tells us how well the parameters $\Theta$ explain the data $\mathcal{D}$ according to model $\mathcal{M}$:
\begin{eqnarray}\label{eq.likelihood}
    \mathcal{L}=-\frac{1}{2}\Sigma_{i}\left(\frac{\theta_{obs.}^i-\theta_{th.}^i}{\sigma_i}\right)^2\,,
\end{eqnarray}
in which $\theta_{obs.}$ is the observed value  of the angular shadow size of the BHs, $\theta_{th.}$ is the theoretical value (\ref{eq.angular diameter})  of the angular shadow size of the BHs.

\begin{figure*}[ht!]\centering
\includegraphics[width=0.45\textwidth]{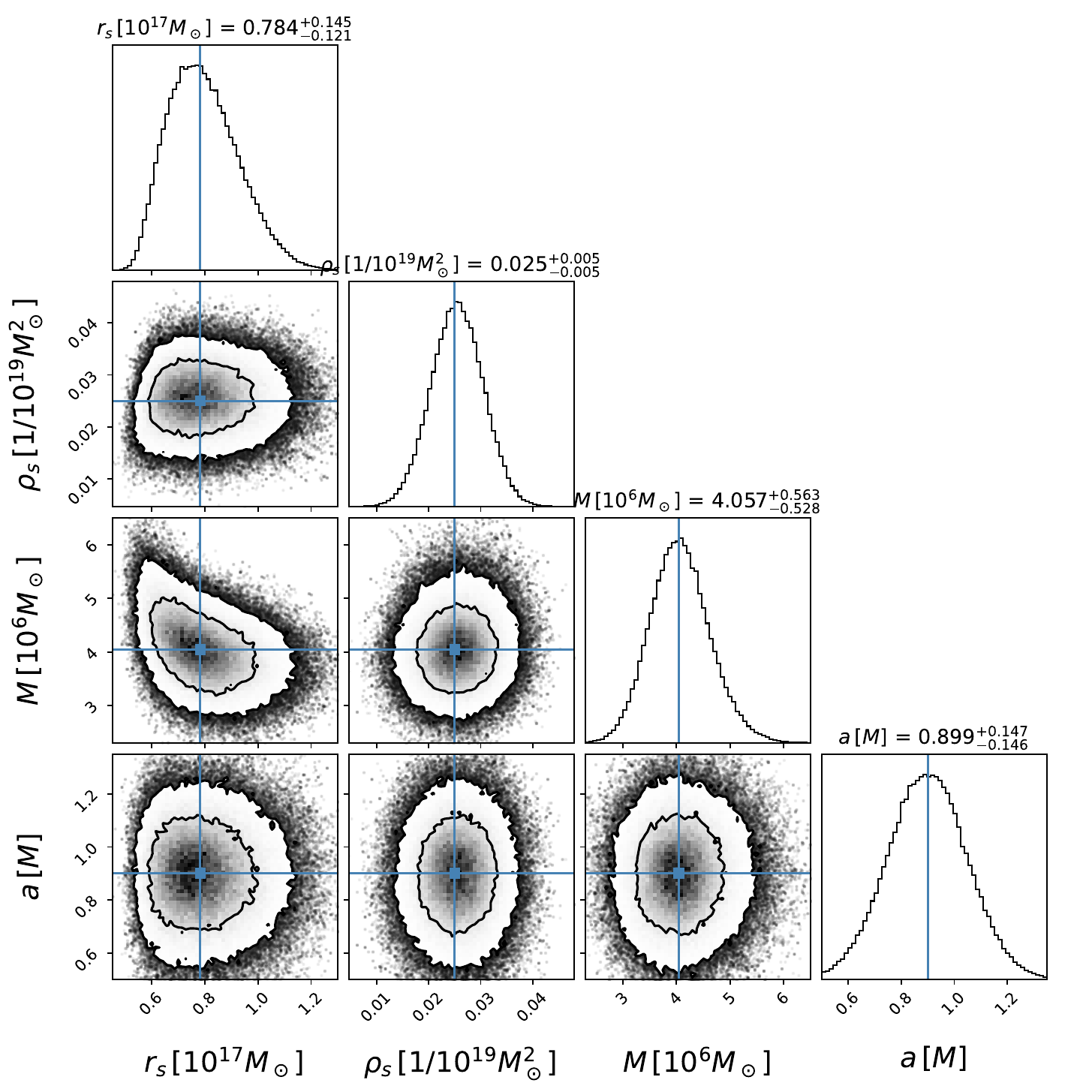}
\includegraphics[width=0.45\textwidth]{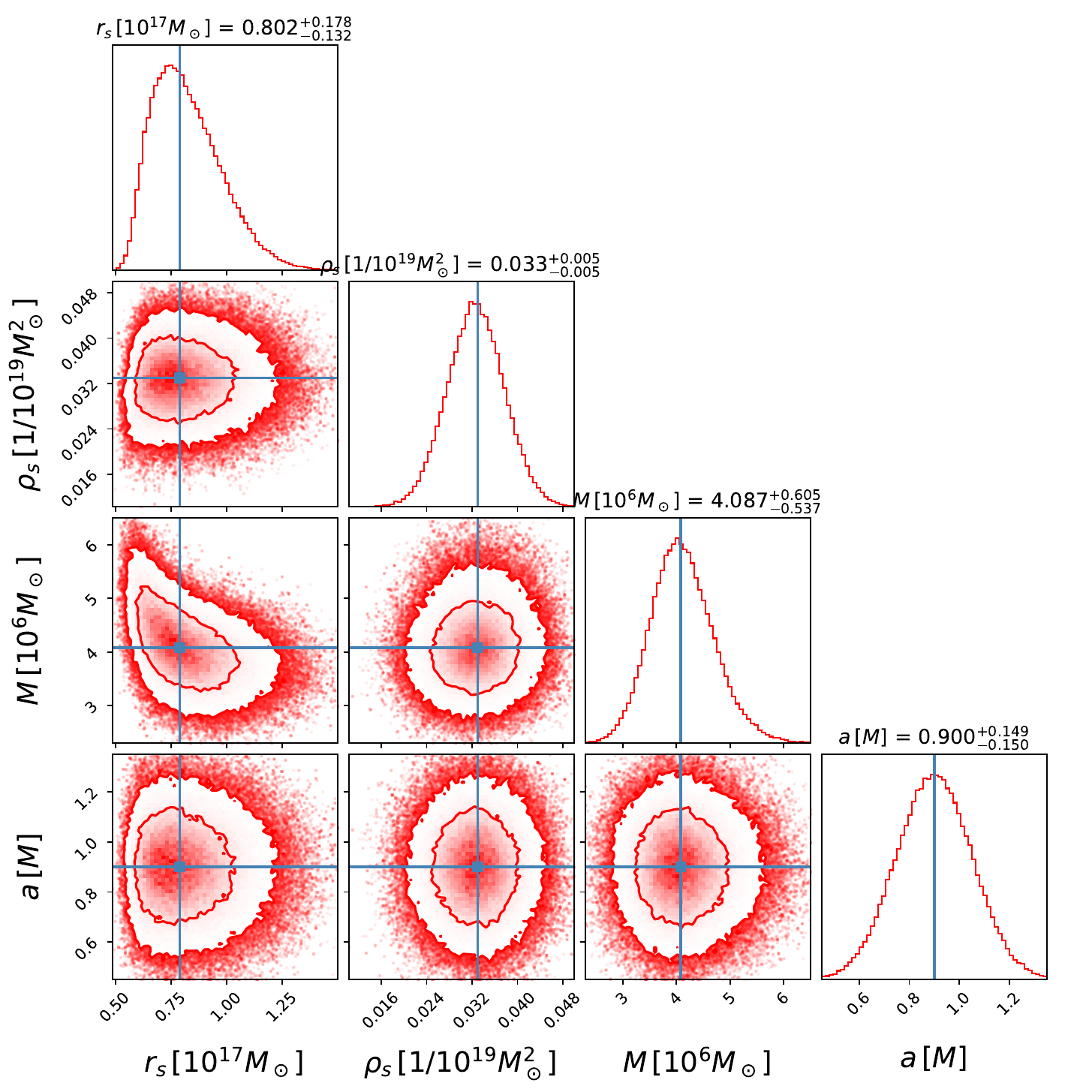}
\includegraphics[width=0.45\textwidth]{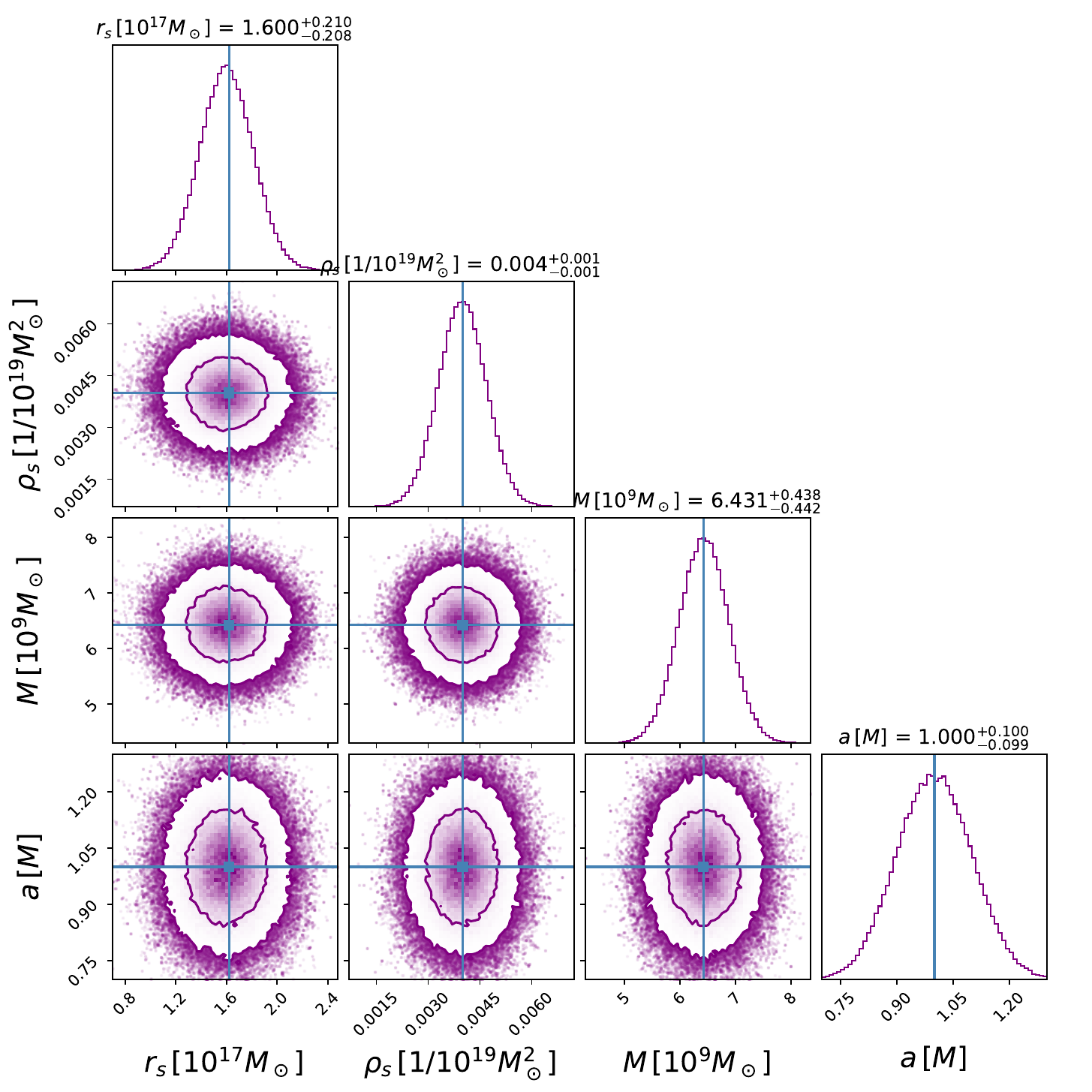}

\caption{MCMC for Sgr A* from EHT (top left), Sgr A* from Gravity (top right), M 87* from EHT (bottom).  
\label{Fig.MCMC}}
\end{figure*}

Then we set the priors  to be Gaussian priors, with boundary condition:
 \begin{eqnarray}\label{eq.Gaussian}
\pi(\Theta_i)\sim\exp{\left[\left(\frac{\Theta_i-\Theta_{0,i}}{\sigma_i}\right)^2\right]},
\end{eqnarray}
in which $\sigma_i$ is the standard deviation.
Subsequently, numerical values for the priors  of the parameters of rotating BH in DM halo used in our analysis is given in Table (\ref{Table 1}).

\begin{table}[h!]
\centering
\resizebox{.5\textwidth}{!}{
\begin{tabular}{|c|cc|cc|cc|cc|cc|}
\hline
& \multicolumn{2}{c|}{$Sgr A^*$ from EHT} & \multicolumn{2}{c|}{$Sgr A^*$ from Gravity} & \multicolumn{2}{c|}{$M 87^*$ from EHT}  \\
 & $\mu$ & $\sigma$ & $\mu$ & $\sigma$ & $\mu$ & $\sigma$ \\
\hline
$M \ [\times 10^6M_\odot]$ & 4.25 & 0.6 & 4.1 & 0.6 & 6300 & 250  \\
$r_s [10^{17}M_\odot]$ & 0.8 & 0.15 & 0.8 & 0.2 & 1.6 & 0.21  \\
$\rho_s[1/10^{19}M_\odot^2]$ & 0.255 & 0.005 & 0.0326 & 0.005 & 0.004 & 0.0007 \\
$a[M]$ & 0.9 & 0.15 & 0.9 & 0.16 & 1 & 0.1 \\
\hline
\end{tabular}
}
\caption{ Gaussian prior on   rotating BHs in Dehnen-type DM halo  from angular shadow size constraints.}
\label{Table 1}
\end{table}

Following, we constrain DM halo parameters $\rho_s$, $r_s$ and BH mass $M$, BH spin $a$ employing MCMC analysis and show visually in Fig.(\ref{Fig.MCMC}).
Also, to be more informative we give numerical values of best-fit values for the parameters in Table(\ref{Table 2}).

\begin{table}[ht!]
    \centering
    \resizebox{.47\textwidth}{!}{
    \begin{tabular}{|l|c|c|c|r|}
     \hline
        & $Sgr A^*$ from EHT & $Sgr A^*$ from Gravity & $M87^*$ from EHT \\
    \hline
      $M(\times 10^6M_\odot)$   & $4.057^{+0.563}_{-0.528}$    & $4.087^{+0.605}_{-0.537}$ & $6.431^{+0.438}_{-0.442}\times10^3$\\
      & & &\\
      $r_s[10^{16}M_\odot]$  & $0.784^{+0.145}_{-0.121}$   & $0.802^{+0.178}_{-0.132}$ & $1.6^{+0.210}_{-0.208}$\\
      & & &\\
      $\rho_s[1/10^{19}M_\odot^2]$   & $0.025^{+0.005}_{-0.005}$   & $0.033^{+0.005}_{-0.005}$ & $0.004^{+0.001}_{-0.001}$\\
      & & &\\
      $a[M]$   & $0.899^{+0.147}_{-0.146}$   & $0.9^{+0.149}_{-0.150}$ & $1^{+0.100}_{-0.099}$\\
      
   \hline
    \end{tabular}}
    \caption{Best-fit rotating BHs in DM halo parameters derived from angular shadow size.}
    \label{Table 2}
\end{table}
Our derived best-fit parameters for the dark matter halo show well agreement with values reported in prior studies (see for example, \cite{Cautun:2019eaf}) as:
\begin{eqnarray}\label{eq.comparision}
  \rho_s\sim \frac{10^{-2}}{(10^{19}M_\odot)^2} \frac{c^6}{G^3}\sim10^{-24}g/cm^{3}\,,\\\nonumber
  r_s\sim10^{17}M_\odot\frac{G}{c^2}\sim 10^{22}cm\sim 10\text{kpc}\,.
\end{eqnarray}
 
\section{Thin accretion disk}\label{Sec.VI}

Now we aim to get a simulation of a geometrically thin infinite accretion disk in a rotating BH surrounded by a DM halo spacetime (\ref{eq.metric components}). To achieve our intention, we have modified the open source Gyoto ray-tracing code (\cite{2011CQGra..28v5011V}) to adapt the code to our obtained spacetime (\ref{eq.metric components}). 

Finding innermost stable circular orbits (ISCO) is an important step to analyze the thin accretion disk of the rotating BH in DM halo (\cite{1996CQGra..13..393M}). To do it we start with expressing the effective potential of the massive particle in the vicinity of a rotating BH surrounded by Dehnen-type DM halo 
(\cite{Khan:2024jez}):

\begin{eqnarray}\label{eq.effective}
    V_{m.eff.}=\frac{g_{t\phi}}{g_{\phi\phi}}l+\sqrt{\left(\frac{g_{t\phi}^2}{g_{\phi\phi}}-g_{tt}\right)\left(1+g^{\phi\phi}l^2\right)},
\end{eqnarray}

where $l=\frac{L}{m}$ is the specific angular momentum of the test particle. Then the ISCO position can be found using the condition:
\begin{eqnarray}\label{eq.ISCO}
 V_{m.eff.}=\frac{\partial V_{m.eff}}{\partial r}=\frac{\partial^2 V_{m.eff}}{\partial r^2}=0. 
\end{eqnarray}

\begin{figure*}[ht!]\centering
\includegraphics[width=0.45\textwidth]{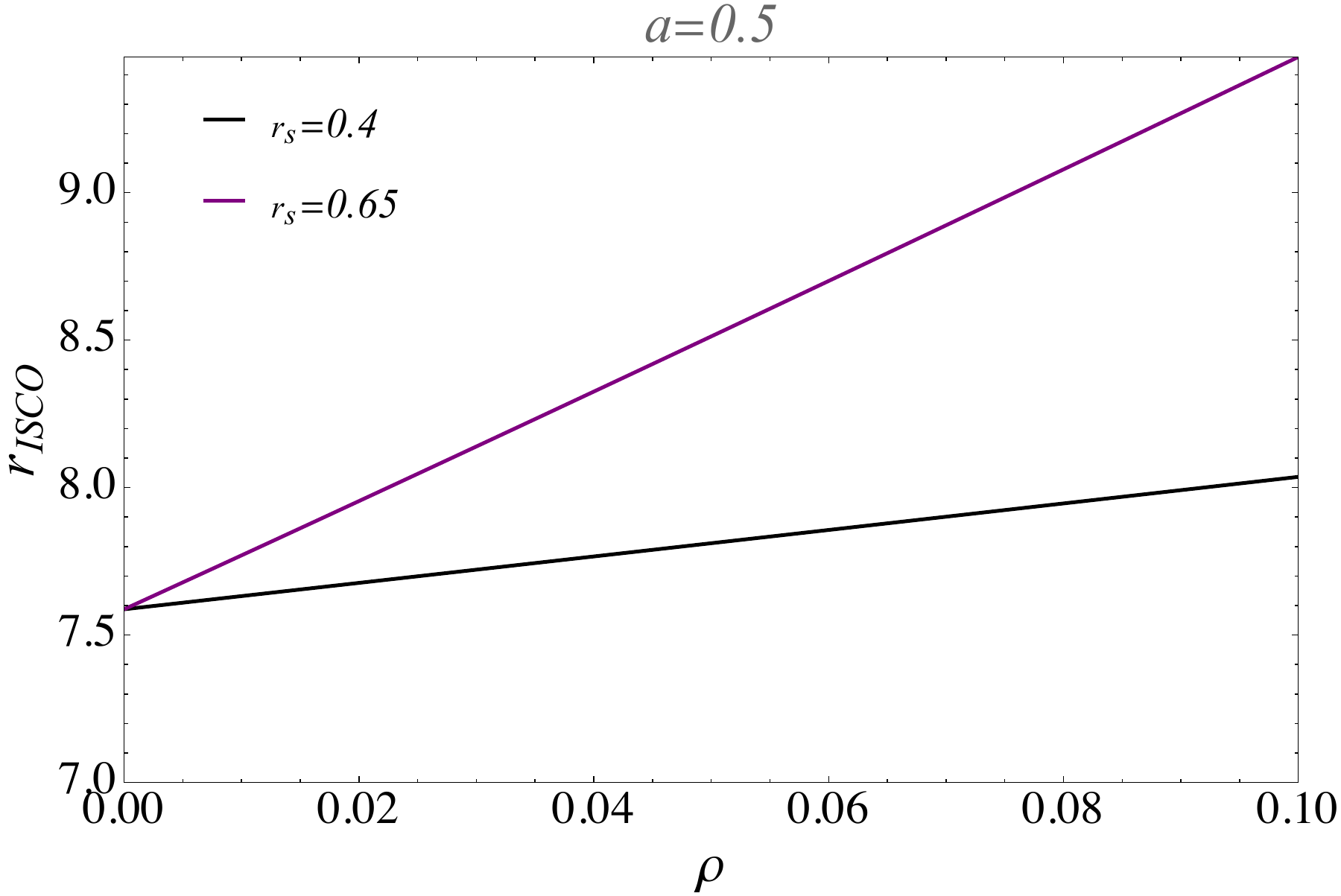}
\includegraphics[width=0.45\textwidth]{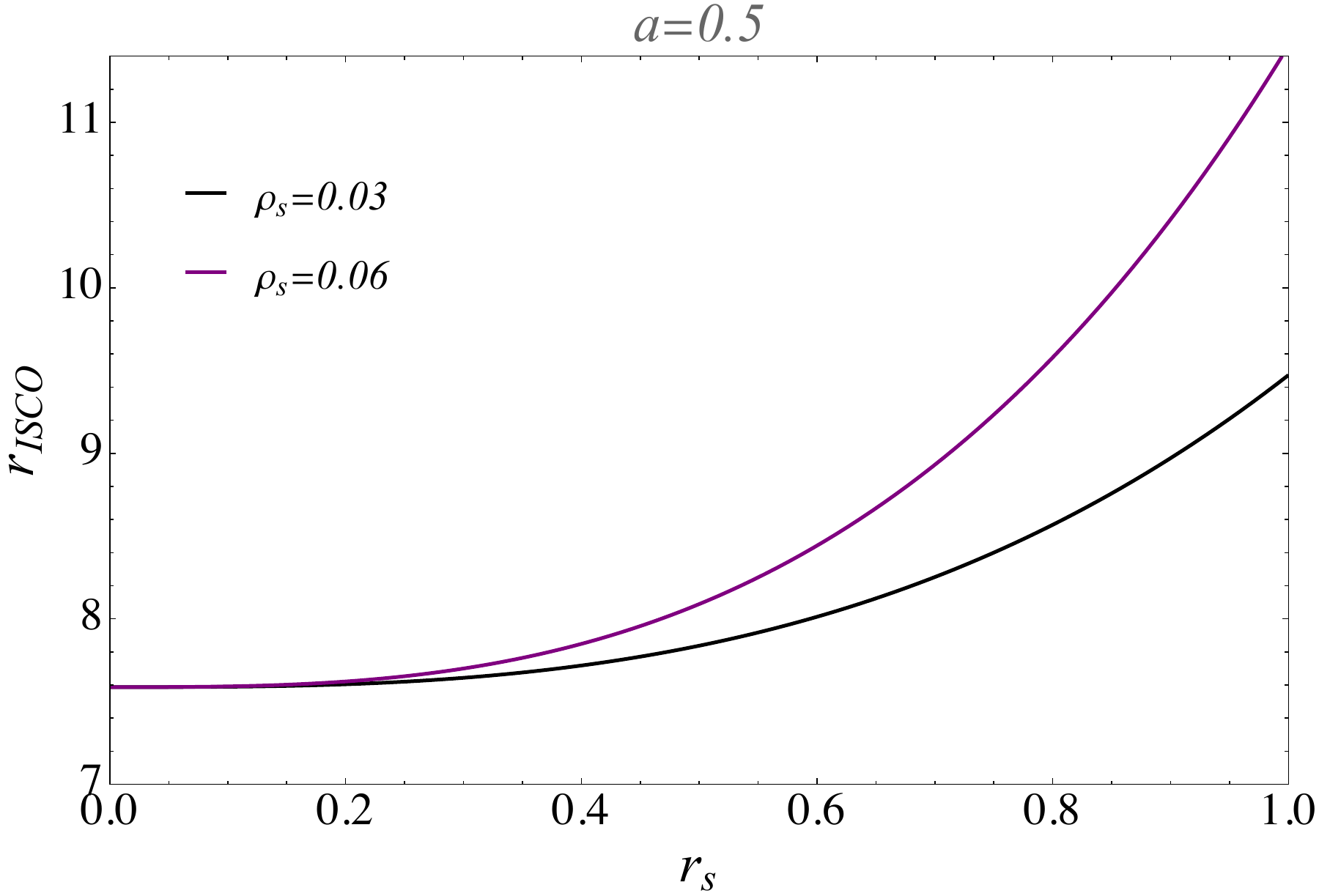}
\caption{The dependence of the  ISCO radius on the parameters of the DM halo $\rho_s$ and $r_s$ for different cases.
\label{Fig.ISCO}}
\end{figure*}

Then, solving Eq.(\ref{eq.ISCO}) numerically, we have plotted how the ISCO radius depends on the DM halo parameters $\rho_s$ and $r_s$ for different cases in Fig.\ref{Fig.ISCO}. It is obvious from Fig.\ref{Fig.ISCO} that increasing both DM halo parameters $\rho_s$ and $r_s$ leads to the enlargement of the value of the ISCO radius $r_{ISCO}$ since the DM halo causes the mass of the BH-DM system.

The flux of electromagnetic radiation can be expressed as (\cite{Uktamov:2024ckf,1Alloqulov2024,Alloqulov2024,Alloqulov2025,1Alloqulov2025}):
\begin{eqnarray}\label{eq.Flux}
  \mathcal{F}(r)=-\frac{\Dot{M_0}}{4\pi}\frac{\Omega_{,r}}{\sqrt{-g}\left(E-\Omega L\right)^2}\int_{r_{ISCO}}^r\left(E-\Omega L\right)L_{,r}dr,  
\end{eqnarray}
here $\Dot{M_0}$ is the accretion rate, $g$
 is the determinant of the three-dimensional subspace.

\begin{figure*}[ht!]\centering
\includegraphics[width=0.45\textwidth]{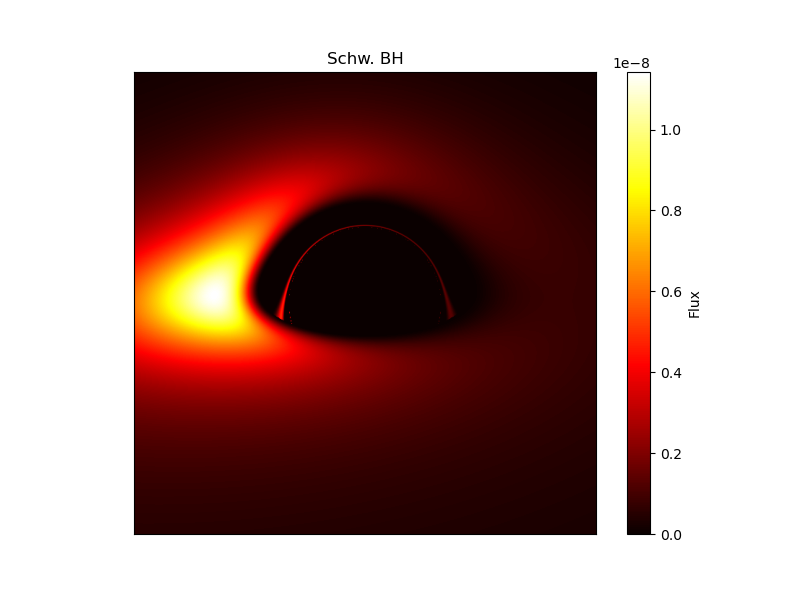}
\includegraphics[width=0.45\textwidth]{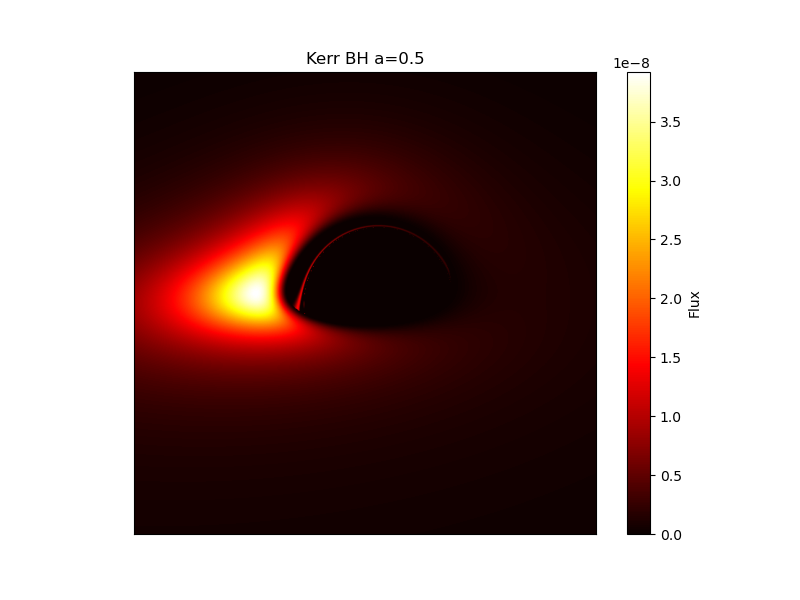}
\includegraphics[width=0.45\textwidth]{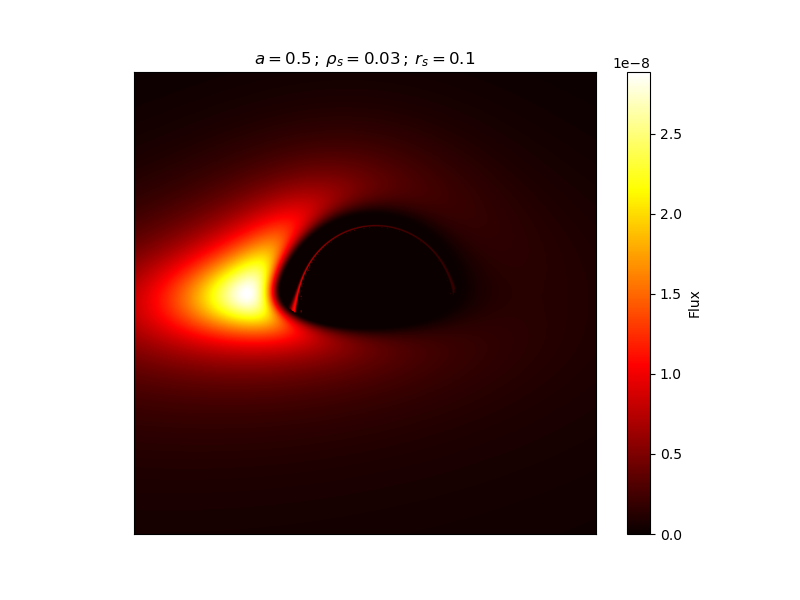}
\includegraphics[width=0.45\textwidth]{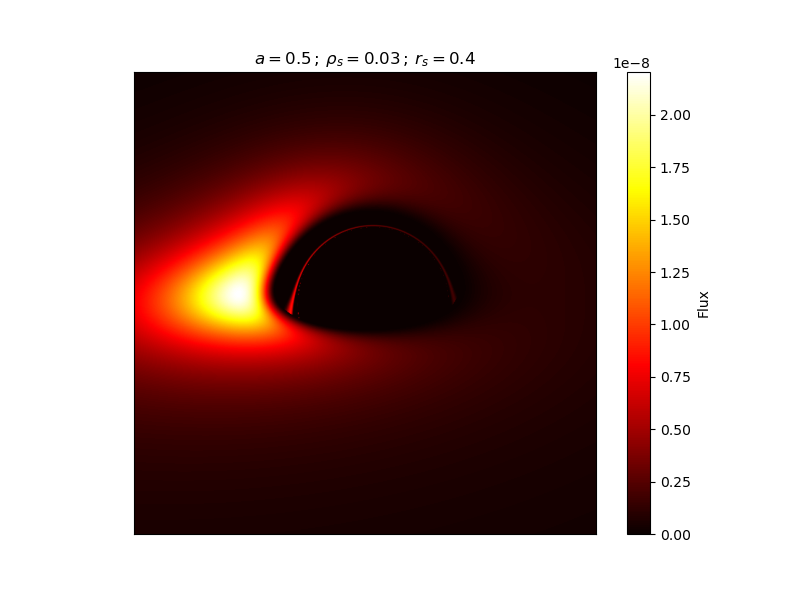}
\includegraphics[width=0.45\textwidth]{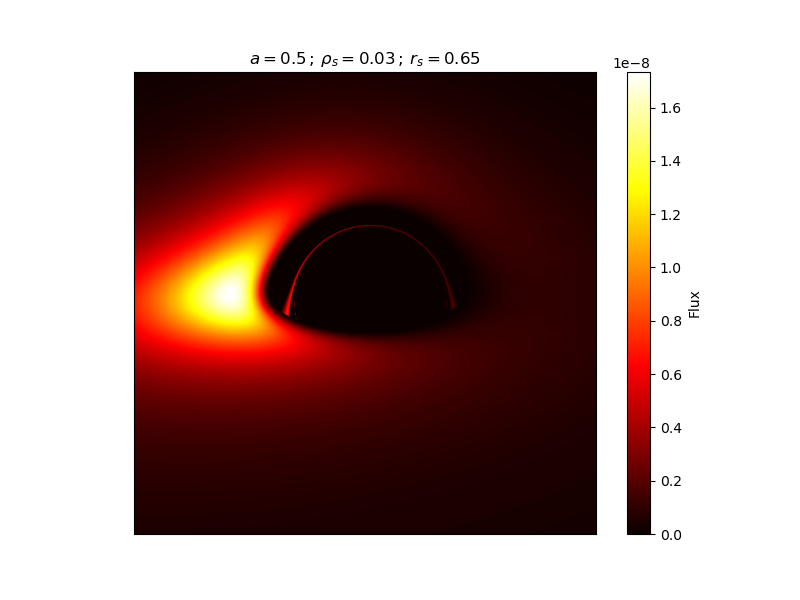}
\includegraphics[width=0.45\textwidth]{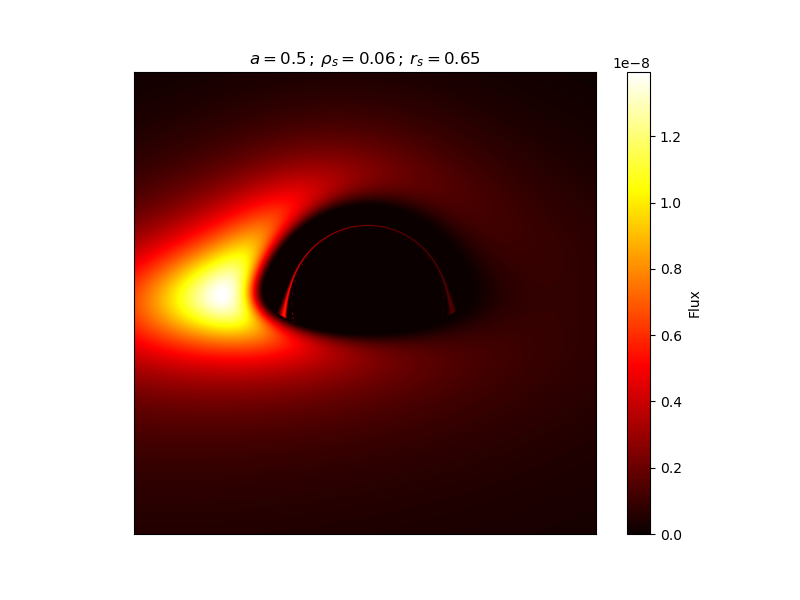}
\caption{The images of the thin accretion disk in Schwarzschild BH, Kerr BH and rotating BH immersed in Dehnen-type DM halo spacetime with the observational angle $\theta_0=\frac{\pi}{10}$.
\label{Fig.flux}}
\end{figure*}

Then, in Fig.\ref{Fig.flux} we have plotted the image thin accretion disk in a rotating BH surrounded by a Dehnen-type DM halo for different values of the DM parameters $\rho_s$ and $r_s$ for the angle of inclination $\theta_0=\frac{\pi}{10}$. One can notice by comparing images of the thin accretion disks of the obtained rotating BHs with thin accretion disks of Kerr and Schwarzschild BHs in Fig.\ref{Fig.flux}, that increasing both DM halo parameters $\rho_s$ and $r_s$ causes decreasing the value of the radiation flux as these parameters $\rho_s$, $r_s$ lead to growing the mass of the BH-DM system, so extending gravity around rotating BH in the DM halo.

\section{Conclusion}\label{Sec.conclusion}

The Event Horizon Telescope (EHT) observations of the shadows of supermassive black holes, like Sgr A* at the Milky Way’s center and M87* in the nearby Messier 87 galaxy, provide an exceptional and natural laboratory for studying the gravitational properties of spacetime geometry around black holes, allowing tests in both strong and weak gravitational field regimes. In this work, we have generated a rotating BH space-time in a dark matter halo by applying the NJA method to the static spacetime metric and the components of the obtained metric tensor are provided in (\ref{eq.metric components}). Then the dependence of the horizon radius on the BH spin $a$ is shown in Fig.~\ref{Fig.horizon} for different values of the characteristic density $\rho_s$ and the characteristic scale factor $r_s$. It is shown that increasing spacetime parameters $r_s$ and $\rho_s$ causes an enlargement of the horizon radius of the rotating BH in a dark matter halo. 

We have also studied the shadow cast by the spinning axially symmetric black hole surrounded by a Dehnen-type dark matter halo. Using the Hamilton-Jacobi equation, we found photon motion in the vicinity of the rotating BH in a dark matter halo, which enabled us to find the photon sphere $r_{ph}$ and the shadow radius. The results obtained showed that the presence of the dark-matter halo causes an enlarged shadow as presented in Fig.~\ref{Fig.shadow}.

Additionally, we have analyzed the photon sphere and shadow radius of the rotating BH in a dark matter halo by giving numerical values of the photon sphere and shadow radius in Fig.~\ref{Fig.shadow radius}.

To strengthen our pure theoretical analysis, we got constraints for BH parameters $\rho_s$ and $r_s$ from EHT observational data for spacetime parameters of supermassive black holes SgrA$*$ and M87$*$ in Sec.~\ref{sec.EHT}. Our provided analysis for the angular diameter $\Theta_d$ demonstrates that the characteristic density $\rho_s$ and the characteristic scale factor $r_s$ of the rotating BH in the Dehnen-type dark matter halo should be relatively high to fit the EHT observed data of the SMBHs images (look to~Fig.\ref{Fig.constrain}).

One may conclude from the performed study that the influence of a Dehnen-type dark matter halo in the environment of supermassive black holes Sgr A$*$ and M87$*$ can not be excluded. To be more precise, we perform MCMC analysis to constrain DM parameters $\rho_s$, $r_s$ and BH mass $M$, BH spin $a$ and it is presented that obtained best-fit values for the DM parameters $\rho_s$ and $r_s$ are well agreement with values showed in prior studies.  

Finally, in Sec.~\ref{Sec.VI} we have analyzed thin accretion disk in a new BH surrounded by a DM halo and give an image of the thin accretion disk in Fig.\ref{Fig.flux}. It was found that increasing DM parameters $\rho_s$ and $r_s$ causes a growing mass of the BH-DM system so leading to a decrease in the value of the electromagnetic radiation.

\section{Acknowledgement}
S.S. is supported by the National Natural Science Foundation of China under Grant No. W2433018.

\bibliographystyle{apsrev4-1}  
\bibliography{A_shadow/main}

\begin{thebibliography}{52}%
\makeatletter
\providecommand \@ifxundefined [1]{%
 \@ifx{#1\undefined}
}%
\providecommand \@ifnum [1]{%
 \ifnum #1\expandafter \@firstoftwo
 \else \expandafter \@secondoftwo
 \fi
}%
\providecommand \@ifx [1]{%
 \ifx #1\expandafter \@firstoftwo
 \else \expandafter \@secondoftwo
 \fi
}%
\providecommand \natexlab [1]{#1}%
\providecommand \enquote  [1]{``#1''}%
\providecommand \bibnamefont  [1]{#1}%
\providecommand \bibfnamefont [1]{#1}%
\providecommand \citenamefont [1]{#1}%
\providecommand \href@noop [0]{\@secondoftwo}%
\providecommand \href [0]{\begingroup \@sanitize@url \@href}%
\providecommand \@href[1]{\@@startlink{#1}\@@href}%
\providecommand \@@href[1]{\endgroup#1\@@endlink}%
\providecommand \@sanitize@url [0]{\catcode `\\12\catcode `\$12\catcode `\&12\catcode `\#12\catcode `\^12\catcode `\_12\catcode `\%12\relax}%
\providecommand \@@startlink[1]{}%
\providecommand \@@endlink[0]{}%
\providecommand \url  [0]{\begingroup\@sanitize@url \@url }%
\providecommand \@url [1]{\endgroup\@href {#1}{\urlprefix }}%
\providecommand \urlprefix  [0]{URL }%
\providecommand \Eprint [0]{\href }%
\providecommand \doibase [0]{http://dx.doi.org/}%
\providecommand \selectlanguage [0]{\@gobble}%
\providecommand \bibinfo  [0]{\@secondoftwo}%
\providecommand \bibfield  [0]{\@secondoftwo}%
\providecommand \translation [1]{[#1]}%
\providecommand \BibitemOpen [0]{}%
\providecommand \bibitemStop [0]{}%
\providecommand \bibitemNoStop [0]{.\EOS\space}%
\providecommand \EOS [0]{\spacefactor3000\relax}%
\providecommand \BibitemShut  [1]{\csname bibitem#1\endcsname}%
\let\auto@bib@innerbib\@empty
\bibitem [{\citenamefont {Abbott}\ \emph {et~al.}(2016)\citenamefont {Abbott} \emph {et~al.}}]{LIGOScientific:2016aoc}%
  \BibitemOpen
  \bibfield  {author} {\bibinfo {author} {\bibfnamefont {B.~P.}\ \bibnamefont {Abbott}} \emph {et~al.} (\bibinfo {collaboration} {LIGO Scientific, Virgo}),\ }\href {\doibase 10.1103/PhysRevLett.116.061102} {\bibfield  {journal} {\bibinfo  {journal} {Phys. Rev. Lett.}\ }\textbf {\bibinfo {volume} {116}},\ \bibinfo {pages} {061102} (\bibinfo {year} {2016})},\ \Eprint {http://arxiv.org/abs/1602.03837} {arXiv:1602.03837 [gr-qc]} \BibitemShut {NoStop}%
\bibitem [{\citenamefont {{Abbott}}\ and\ \citenamefont {{Abbott}}(2016)}]{2016PhRvL.116x1102A}%
  \BibitemOpen
  \bibfield  {author} {\bibinfo {author} {\bibfnamefont {B.~P.}\ \bibnamefont {{Abbott}}}\ and\ \bibinfo {author} {\bibfnamefont {J.~D.~E.}\ \bibnamefont {{Abbott}}},\ }\href {\doibase 10.1103/PhysRevLett.116.241102} {\bibfield  {journal} {\bibinfo  {journal} {Phys. Rev. Lett.}\ }\textbf {\bibinfo {volume} {116}},\ \bibinfo {eid} {241102} (\bibinfo {year} {2016})},\ \Eprint {http://arxiv.org/abs/1602.03840} {arXiv:1602.03840 [gr-qc]} \BibitemShut {NoStop}%
\bibitem [{\citenamefont {{Iocco}}\ \emph {et~al.}(2015)\citenamefont {{Iocco}}, \citenamefont {{Pato}},\ and\ \citenamefont {{Bertone}}}]{2015NatPh..11..245I}%
  \BibitemOpen
  \bibfield  {author} {\bibinfo {author} {\bibfnamefont {F.}~\bibnamefont {{Iocco}}}, \bibinfo {author} {\bibfnamefont {M.}~\bibnamefont {{Pato}}}, \ and\ \bibinfo {author} {\bibfnamefont {G.}~\bibnamefont {{Bertone}}},\ }\href {\doibase 10.1038/nphys3237} {\bibfield  {journal} {\bibinfo  {journal} {Nat. Phys.}\ }\textbf {\bibinfo {volume} {11}},\ \bibinfo {pages} {245} (\bibinfo {year} {2015})},\ \Eprint {http://arxiv.org/abs/1502.03821} {arXiv:1502.03821 [astro-ph.GA]} \BibitemShut {NoStop}%
\bibitem [{\citenamefont {{Bertone}}\ and\ \citenamefont {{Tait}}(2018)}]{2018Natur.562...51B}%
  \BibitemOpen
  \bibfield  {author} {\bibinfo {author} {\bibfnamefont {G.}~\bibnamefont {{Bertone}}}\ and\ \bibinfo {author} {\bibfnamefont {T.~M.~P.}\ \bibnamefont {{Tait}}},\ }\href {\doibase 10.1038/s41586-018-0542-z} {\bibfield  {journal} {\bibinfo  {journal} {Nature}\ }\textbf {\bibinfo {volume} {562}},\ \bibinfo {pages} {51} (\bibinfo {year} {2018})},\ \Eprint {http://arxiv.org/abs/1810.01668} {arXiv:1810.01668 [astro-ph.CO]} \BibitemShut {NoStop}%
\bibitem [{\citenamefont {Graham}\ \emph {et~al.}(2006)\citenamefont {Graham}, \citenamefont {Merritt}, \citenamefont {Moore}, \citenamefont {Diemand},\ and\ \citenamefont {Terzic}}]{Graham:2006ae}%
  \BibitemOpen
  \bibfield  {author} {\bibinfo {author} {\bibfnamefont {A.~W.}\ \bibnamefont {Graham}}, \bibinfo {author} {\bibfnamefont {D.}~\bibnamefont {Merritt}}, \bibinfo {author} {\bibfnamefont {B.}~\bibnamefont {Moore}}, \bibinfo {author} {\bibfnamefont {J.}~\bibnamefont {Diemand}}, \ and\ \bibinfo {author} {\bibfnamefont {B.}~\bibnamefont {Terzic}},\ }\href {\doibase 10.1086/508990} {\bibfield  {journal} {\bibinfo  {journal} {Astron. J.}\ }\textbf {\bibinfo {volume} {132}},\ \bibinfo {pages} {2701} (\bibinfo {year} {2006})},\ \Eprint {http://arxiv.org/abs/astro-ph/0608613} {arXiv:astro-ph/0608613} \BibitemShut {NoStop}%
\bibitem [{\citenamefont {{Navarro}}\ \emph {et~al.}(1996)\citenamefont {{Navarro}}, \citenamefont {{Frenk}},\ and\ \citenamefont {{White}}}]{1996ApJ...462..563N}%
  \BibitemOpen
  \bibfield  {author} {\bibinfo {author} {\bibfnamefont {J.~F.}\ \bibnamefont {{Navarro}}}, \bibinfo {author} {\bibfnamefont {C.~S.}\ \bibnamefont {{Frenk}}}, \ and\ \bibinfo {author} {\bibfnamefont {S.~D.~M.}\ \bibnamefont {{White}}},\ }\href {\doibase 10.1086/177173} {\bibfield  {journal} {\bibinfo  {journal} {Astrophys. J.}\ }\textbf {\bibinfo {volume} {462}},\ \bibinfo {pages} {563} (\bibinfo {year} {1996})},\ \Eprint {http://arxiv.org/abs/astro-ph/9508025} {arXiv:astro-ph/9508025 [astro-ph]} \BibitemShut {NoStop}%
\bibitem [{\citenamefont {{Burkert}}(1995)}]{1995ApJ...447L..25B}%
  \BibitemOpen
  \bibfield  {author} {\bibinfo {author} {\bibfnamefont {A.}~\bibnamefont {{Burkert}}},\ }\href {\doibase 10.1086/309560} {\bibfield  {journal} {\bibinfo  {journal} {Astrophys. J. Lett.}\ }\textbf {\bibinfo {volume} {447}},\ \bibinfo {pages} {L25} (\bibinfo {year} {1995})},\ \Eprint {http://arxiv.org/abs/astro-ph/9504041} {arXiv:astro-ph/9504041 [astro-ph]} \BibitemShut {NoStop}%
\bibitem [{\citenamefont {{Dehnen}}(1993)}]{1993MNRAS.265..250D}%
  \BibitemOpen
  \bibfield  {author} {\bibinfo {author} {\bibfnamefont {W.}~\bibnamefont {{Dehnen}}},\ }\href {\doibase 10.1093/mnras/265.1.250} {\bibfield  {journal} {\bibinfo  {journal} {Mon. Not. R. Astron. Soc.}\ }\textbf {\bibinfo {volume} {265}},\ \bibinfo {pages} {250} (\bibinfo {year} {1993})}\BibitemShut {NoStop}%
\bibitem [{\citenamefont {{Shukirgaliyev}}\ \emph {et~al.}(2021)\citenamefont {{Shukirgaliyev}}, \citenamefont {{Otebay}}, \citenamefont {{Sobolenko}}, \citenamefont {{Ishchenko}}, \citenamefont {{Borodina}}, \citenamefont {{Panamarev}}, \citenamefont {{Myrzakul}}, \citenamefont {{Kalambay}}, \citenamefont {{Naurzbayeva}}, \citenamefont {{Abdikamalov}}, \citenamefont {{Polyachenko}}, \citenamefont {{Banerjee}}, \citenamefont {{Berczik}}, \citenamefont {{Spurzem}},\ and\ \citenamefont {{Just}}}]{2021A&A...654A..53S}%
  \BibitemOpen
  \bibfield  {author} {\bibinfo {author} {\bibfnamefont {B.}~\bibnamefont {{Shukirgaliyev}}}, \bibinfo {author} {\bibfnamefont {A.}~\bibnamefont {{Otebay}}}, \bibinfo {author} {\bibfnamefont {M.}~\bibnamefont {{Sobolenko}}}, \bibinfo {author} {\bibfnamefont {M.}~\bibnamefont {{Ishchenko}}}, \bibinfo {author} {\bibfnamefont {O.}~\bibnamefont {{Borodina}}}, \bibinfo {author} {\bibfnamefont {T.}~\bibnamefont {{Panamarev}}}, \bibinfo {author} {\bibfnamefont {S.}~\bibnamefont {{Myrzakul}}}, \bibinfo {author} {\bibfnamefont {M.}~\bibnamefont {{Kalambay}}}, \bibinfo {author} {\bibfnamefont {A.}~\bibnamefont {{Naurzbayeva}}}, \bibinfo {author} {\bibfnamefont {E.}~\bibnamefont {{Abdikamalov}}}, \bibinfo {author} {\bibfnamefont {E.}~\bibnamefont {{Polyachenko}}}, \bibinfo {author} {\bibfnamefont {S.}~\bibnamefont {{Banerjee}}}, \bibinfo {author} {\bibfnamefont {P.}~\bibnamefont {{Berczik}}}, \bibinfo {author} {\bibfnamefont {R.}~\bibnamefont {{Spurzem}}}, \ and\ \bibinfo {author} {\bibfnamefont {A.}~\bibnamefont
  {{Just}}},\ }\href {\doibase 10.1051/0004-6361/202141299} {\bibfield  {journal} {\bibinfo  {journal} {Astron. Astrophys.}\ }\textbf {\bibinfo {volume} {654}},\ \bibinfo {eid} {A53} (\bibinfo {year} {2021})},\ \Eprint {http://arxiv.org/abs/2105.09510} {arXiv:2105.09510 [astro-ph.GA]} \BibitemShut {NoStop}%
\bibitem [{\citenamefont {Al-Badawi}\ \emph {et~al.}(2025)\citenamefont {Al-Badawi}, \citenamefont {Shaymatov},\ and\ \citenamefont {Sekhmani}}]{Al-Badawi:2024asn}%
  \BibitemOpen
  \bibfield  {author} {\bibinfo {author} {\bibfnamefont {A.}~\bibnamefont {Al-Badawi}}, \bibinfo {author} {\bibfnamefont {S.}~\bibnamefont {Shaymatov}}, \ and\ \bibinfo {author} {\bibfnamefont {Y.}~\bibnamefont {Sekhmani}},\ }\href {\doibase 10.1088/1475-7516/2025/02/014} {\bibfield  {journal} {\bibinfo  {journal} {JCAP}\ }\textbf {\bibinfo {volume} {02}},\ \bibinfo {pages} {014} (\bibinfo {year} {2025})},\ \Eprint {http://arxiv.org/abs/2411.01145} {arXiv:2411.01145 [gr-qc]} \BibitemShut {NoStop}%
\bibitem [{\citenamefont {{Maeda}}\ \emph {et~al.}(2025)\citenamefont {{Maeda}}, \citenamefont {{Cardoso}},\ and\ \citenamefont {{Wang}}}]{2025PhRvD.111d4060M}%
  \BibitemOpen
  \bibfield  {author} {\bibinfo {author} {\bibfnamefont {K.-i.}\ \bibnamefont {{Maeda}}}, \bibinfo {author} {\bibfnamefont {V.}~\bibnamefont {{Cardoso}}}, \ and\ \bibinfo {author} {\bibfnamefont {A.}~\bibnamefont {{Wang}}},\ }\href {\doibase 10.1103/PhysRevD.111.044060} {\bibfield  {journal} {\bibinfo  {journal} {Phys. Rev. D}\ }\textbf {\bibinfo {volume} {111}},\ \bibinfo {eid} {044060} (\bibinfo {year} {2025})},\ \Eprint {http://arxiv.org/abs/2410.04175} {arXiv:2410.04175 [gr-qc]} \BibitemShut {NoStop}%
\bibitem [{\citenamefont {{Shen}}\ \emph {et~al.}(2025)\citenamefont {{Shen}}, \citenamefont {{Wang}},\ and\ \citenamefont {{Yin}}}]{2025PhLB..86239300S}%
  \BibitemOpen
  \bibfield  {author} {\bibinfo {author} {\bibfnamefont {Z.}~\bibnamefont {{Shen}}}, \bibinfo {author} {\bibfnamefont {A.}~\bibnamefont {{Wang}}}, \ and\ \bibinfo {author} {\bibfnamefont {S.}~\bibnamefont {{Yin}}},\ }\href {\doibase 10.1016/j.physletb.2025.139300} {\bibfield  {journal} {\bibinfo  {journal} {Phys. Lett. B}\ }\textbf {\bibinfo {volume} {862}},\ \bibinfo {eid} {139300} (\bibinfo {year} {2025})},\ \Eprint {http://arxiv.org/abs/2408.05417} {arXiv:2408.05417 [gr-qc]} \BibitemShut {NoStop}%
\bibitem [{\citenamefont {{Shen}}\ \emph {et~al.}(2024)\citenamefont {{Shen}}, \citenamefont {{Wang}}, \citenamefont {{Gong}},\ and\ \citenamefont {{Yin}}}]{2024PhLB..85538797S}%
  \BibitemOpen
  \bibfield  {author} {\bibinfo {author} {\bibfnamefont {Z.}~\bibnamefont {{Shen}}}, \bibinfo {author} {\bibfnamefont {A.}~\bibnamefont {{Wang}}}, \bibinfo {author} {\bibfnamefont {Y.}~\bibnamefont {{Gong}}}, \ and\ \bibinfo {author} {\bibfnamefont {S.}~\bibnamefont {{Yin}}},\ }\href {\doibase 10.1016/j.physletb.2024.138797} {\bibfield  {journal} {\bibinfo  {journal} {Phys. Lett. B}\ }\textbf {\bibinfo {volume} {855}},\ \bibinfo {eid} {138797} (\bibinfo {year} {2024})},\ \Eprint {http://arxiv.org/abs/2311.12259} {arXiv:2311.12259 [gr-qc]} \BibitemShut {NoStop}%
\bibitem [{\citenamefont {Xamidov}\ \emph {et~al.}(2025)\citenamefont {Xamidov}, \citenamefont {Shaymatov}, \citenamefont {Wu},\ and\ \citenamefont {Zhu}}]{Xamidov:2025prl}%
  \BibitemOpen
  \bibfield  {author} {\bibinfo {author} {\bibfnamefont {T.}~\bibnamefont {Xamidov}}, \bibinfo {author} {\bibfnamefont {S.}~\bibnamefont {Shaymatov}}, \bibinfo {author} {\bibfnamefont {Q.}~\bibnamefont {Wu}}, \ and\ \bibinfo {author} {\bibfnamefont {T.}~\bibnamefont {Zhu}},\ }\href {\doibase 10.1140/epjc/s10052-025-14912-5} {\bibfield  {journal} {\bibinfo  {journal} {Eur. Phys. J. C}\ }\textbf {\bibinfo {volume} {85}},\ \bibinfo {pages} {1193} (\bibinfo {year} {2025})},\ \Eprint {http://arxiv.org/abs/2507.13147} {arXiv:2507.13147 [gr-qc]} \BibitemShut {NoStop}%
\bibitem [{\citenamefont {{Alloqulov}}\ \emph {et~al.}(2025)\citenamefont {{Alloqulov}}, \citenamefont {{Xamidov}}, \citenamefont {{Shaymatov}},\ and\ \citenamefont {{Ahmedov}}}]{2025arXiv250405236A}%
  \BibitemOpen
  \bibfield  {author} {\bibinfo {author} {\bibfnamefont {M.}~\bibnamefont {{Alloqulov}}}, \bibinfo {author} {\bibfnamefont {T.}~\bibnamefont {{Xamidov}}}, \bibinfo {author} {\bibfnamefont {S.}~\bibnamefont {{Shaymatov}}}, \ and\ \bibinfo {author} {\bibfnamefont {B.}~\bibnamefont {{Ahmedov}}},\ }\href {\doibase 10.48550/arXiv.2504.05236} {\bibfield  {journal} {\bibinfo  {journal} {arXiv e-prints}\ ,\ \bibinfo {eid} {arXiv:2504.05236}} (\bibinfo {year} {2025})},\ \Eprint {http://arxiv.org/abs/2504.05236} {arXiv:2504.05236 [gr-qc]} \BibitemShut {NoStop}%
\bibitem [{\citenamefont {{Gohain}}\ \emph {et~al.}(2024)\citenamefont {{Gohain}}, \citenamefont {{Phukon}},\ and\ \citenamefont {{Bhuyan}}}]{2024PDU....4601683G}%
  \BibitemOpen
  \bibfield  {author} {\bibinfo {author} {\bibfnamefont {M.~M.}\ \bibnamefont {{Gohain}}}, \bibinfo {author} {\bibfnamefont {P.}~\bibnamefont {{Phukon}}}, \ and\ \bibinfo {author} {\bibfnamefont {K.}~\bibnamefont {{Bhuyan}}},\ }\href {\doibase 10.1016/j.dark.2024.101683} {\bibfield  {journal} {\bibinfo  {journal} {Phys. Dark Universe}\ }\textbf {\bibinfo {volume} {46}},\ \bibinfo {eid} {101683} (\bibinfo {year} {2024})},\ \Eprint {http://arxiv.org/abs/2407.02872} {arXiv:2407.02872 [gr-qc]} \BibitemShut {NoStop}%
\bibitem [{\citenamefont {{Xamidov}}\ \emph {et~al.}(2025)\citenamefont {{Xamidov}}, \citenamefont {{Uktamov}}, \citenamefont {{Shaymatov}},\ and\ \citenamefont {{Ahmedov}}}]{2025PDU....4701805X}%
  \BibitemOpen
  \bibfield  {author} {\bibinfo {author} {\bibfnamefont {T.}~\bibnamefont {{Xamidov}}}, \bibinfo {author} {\bibfnamefont {U.}~\bibnamefont {{Uktamov}}}, \bibinfo {author} {\bibfnamefont {S.}~\bibnamefont {{Shaymatov}}}, \ and\ \bibinfo {author} {\bibfnamefont {B.}~\bibnamefont {{Ahmedov}}},\ }\href {\doibase 10.1016/j.dark.2024.101805} {\bibfield  {journal} {\bibinfo  {journal} {Phys. Dark Universe}\ }\textbf {\bibinfo {volume} {47}},\ \bibinfo {eid} {101805} (\bibinfo {year} {2025})}\BibitemShut {NoStop}%
\bibitem [{\citenamefont {Vagnozzi}\ and\ \citenamefont {Visinelli}(2019)}]{PhysRevD.100.024020}%
  \BibitemOpen
  \bibfield  {author} {\bibinfo {author} {\bibfnamefont {S.}~\bibnamefont {Vagnozzi}}\ and\ \bibinfo {author} {\bibfnamefont {L.}~\bibnamefont {Visinelli}},\ }\href {\doibase 10.1103/PhysRevD.100.024020} {\bibfield  {journal} {\bibinfo  {journal} {Phys. Rev. D}\ }\textbf {\bibinfo {volume} {100}},\ \bibinfo {pages} {024020} (\bibinfo {year} {2019})}\BibitemShut {NoStop}%
\bibitem [{\citenamefont {Bambi}\ \emph {et~al.}(2019)\citenamefont {Bambi}, \citenamefont {Freese}, \citenamefont {Vagnozzi},\ and\ \citenamefont {Visinelli}}]{Bambi:2019tjh}%
  \BibitemOpen
  \bibfield  {author} {\bibinfo {author} {\bibfnamefont {C.}~\bibnamefont {Bambi}}, \bibinfo {author} {\bibfnamefont {K.}~\bibnamefont {Freese}}, \bibinfo {author} {\bibfnamefont {S.}~\bibnamefont {Vagnozzi}}, \ and\ \bibinfo {author} {\bibfnamefont {L.}~\bibnamefont {Visinelli}},\ }\href {\doibase 10.1103/PhysRevD.100.044057} {\bibfield  {journal} {\bibinfo  {journal} {Phys. Rev. D}\ }\textbf {\bibinfo {volume} {100}},\ \bibinfo {pages} {044057} (\bibinfo {year} {2019})},\ \Eprint {http://arxiv.org/abs/1904.12983} {arXiv:1904.12983 [gr-qc]} \BibitemShut {NoStop}%
\bibitem [{\citenamefont {{Vagnozzi}}\ \emph {et~al.}(2023)\citenamefont {{Vagnozzi}}, \citenamefont {{Roy}}, \citenamefont {{Tsai}}, \citenamefont {{Visinelli}}, \citenamefont {{Afrin}}, \citenamefont {{Allahyari}}, \citenamefont {{Bambhaniya}}, \citenamefont {{Dey}}, \citenamefont {{Ghosh}}, \citenamefont {{Joshi}}, \citenamefont {{Jusufi}}, \citenamefont {{Khodadi}}, \citenamefont {{Walia}}, \citenamefont {{{\"O}vg{\"u}n}},\ and\ \citenamefont {{Bambi}}}]{2023CQGra..40p5007V}%
  \BibitemOpen
  \bibfield  {author} {\bibinfo {author} {\bibfnamefont {S.}~\bibnamefont {{Vagnozzi}}}, \bibinfo {author} {\bibfnamefont {R.}~\bibnamefont {{Roy}}}, \bibinfo {author} {\bibfnamefont {Y.-D.}\ \bibnamefont {{Tsai}}}, \bibinfo {author} {\bibfnamefont {L.}~\bibnamefont {{Visinelli}}}, \bibinfo {author} {\bibfnamefont {M.}~\bibnamefont {{Afrin}}}, \bibinfo {author} {\bibfnamefont {A.}~\bibnamefont {{Allahyari}}}, \bibinfo {author} {\bibfnamefont {P.}~\bibnamefont {{Bambhaniya}}}, \bibinfo {author} {\bibfnamefont {D.}~\bibnamefont {{Dey}}}, \bibinfo {author} {\bibfnamefont {S.~G.}\ \bibnamefont {{Ghosh}}}, \bibinfo {author} {\bibfnamefont {P.~S.}\ \bibnamefont {{Joshi}}}, \bibinfo {author} {\bibfnamefont {K.}~\bibnamefont {{Jusufi}}}, \bibinfo {author} {\bibfnamefont {M.}~\bibnamefont {{Khodadi}}}, \bibinfo {author} {\bibfnamefont {R.~K.}\ \bibnamefont {{Walia}}}, \bibinfo {author} {\bibfnamefont {A.}~\bibnamefont {{{\"O}vg{\"u}n}}}, \ and\ \bibinfo {author} {\bibfnamefont {C.}~\bibnamefont {{Bambi}}},\ }\href
  {\doibase 10.1088/1361-6382/acd97b} {\bibfield  {journal} {\bibinfo  {journal} {Classical and Quantum Gravity}\ }\textbf {\bibinfo {volume} {40}},\ \bibinfo {eid} {165007} (\bibinfo {year} {2023})},\ \Eprint {http://arxiv.org/abs/2205.07787} {arXiv:2205.07787 [gr-qc]} \BibitemShut {NoStop}%
\bibitem [{\citenamefont {{Akiyama}}\ and\ \citenamefont {et~al. {(Event Horizon Telescope Collaboration)}}(2019)}]{Akiyama19L1}%
  \BibitemOpen
  \bibfield  {author} {\bibinfo {author} {\bibfnamefont {K.}~\bibnamefont {{Akiyama}}}\ and\ \bibinfo {author} {\bibnamefont {et~al. {(Event Horizon Telescope Collaboration)}}},\ }\href {\doibase 10.3847/2041-8213/ab0ec7} {\bibfield  {journal} {\bibinfo  {journal} {Astrophys. J.}\ }\textbf {\bibinfo {volume} {875}},\ \bibinfo {eid} {L1} (\bibinfo {year} {2019})},\ \Eprint {http://arxiv.org/abs/1906.11238} {arXiv:1906.11238 [astro-ph.GA]} \BibitemShut {NoStop}%
\bibitem [{\citenamefont {{Akiyama}}\ and\ \citenamefont {et~al. {(Event Horizon Telescope Collaboration)}}(2022)}]{2022ApJ...930L..12E}%
  \BibitemOpen
  \bibfield  {author} {\bibinfo {author} {\bibfnamefont {K.}~\bibnamefont {{Akiyama}}}\ and\ \bibinfo {author} {\bibnamefont {et~al. {(Event Horizon Telescope Collaboration)}}},\ }\href {\doibase 10.3847/2041-8213/ac6674} {\bibfield  {journal} {\bibinfo  {journal} {Astrophys. J. Lett.}\ }\textbf {\bibinfo {volume} {930}},\ \bibinfo {eid} {L12} (\bibinfo {year} {2022})}\BibitemShut {NoStop}%
\bibitem [{\citenamefont {{Cardoso}}\ and\ \citenamefont {{Pani}}(2019)}]{2019LRR....22....4C}%
  \BibitemOpen
  \bibfield  {author} {\bibinfo {author} {\bibfnamefont {V.}~\bibnamefont {{Cardoso}}}\ and\ \bibinfo {author} {\bibfnamefont {P.}~\bibnamefont {{Pani}}},\ }\href {\doibase 10.1007/s41114-019-0020-4} {\bibfield  {journal} {\bibinfo  {journal} {Living Rev. Relativ.}\ }\textbf {\bibinfo {volume} {22}},\ \bibinfo {eid} {4} (\bibinfo {year} {2019})},\ \Eprint {http://arxiv.org/abs/1904.05363} {arXiv:1904.05363 [gr-qc]} \BibitemShut {NoStop}%
\bibitem [{\citenamefont {{Zahid}}\ \emph {et~al.}(2022)\citenamefont {{Zahid}}, \citenamefont {{Khan}}, \citenamefont {{Ren}},\ and\ \citenamefont {{Rayimbaev}}}]{2022IJMPD..3150058Z}%
  \BibitemOpen
  \bibfield  {author} {\bibinfo {author} {\bibfnamefont {M.}~\bibnamefont {{Zahid}}}, \bibinfo {author} {\bibfnamefont {S.~U.}\ \bibnamefont {{Khan}}}, \bibinfo {author} {\bibfnamefont {J.}~\bibnamefont {{Ren}}}, \ and\ \bibinfo {author} {\bibfnamefont {J.}~\bibnamefont {{Rayimbaev}}},\ }\href {\doibase 10.1142/S0218271822500584} {\bibfield  {journal} {\bibinfo  {journal} {Int. J. Mod. Phys. D}\ }\textbf {\bibinfo {volume} {31}},\ \bibinfo {eid} {2250058} (\bibinfo {year} {2022})}\BibitemShut {NoStop}%
\bibitem [{\citenamefont {{Zubair}}\ \emph {et~al.}(2022)\citenamefont {{Zubair}}, \citenamefont {{Raza}},\ and\ \citenamefont {{Abbas}}}]{2022EPJC...82..948Z}%
  \BibitemOpen
  \bibfield  {author} {\bibinfo {author} {\bibfnamefont {M.}~\bibnamefont {{Zubair}}}, \bibinfo {author} {\bibfnamefont {M.~A.}\ \bibnamefont {{Raza}}}, \ and\ \bibinfo {author} {\bibfnamefont {G.}~\bibnamefont {{Abbas}}},\ }\href {\doibase 10.1140/epjc/s10052-022-10925-6} {\bibfield  {journal} {\bibinfo  {journal} {Eur. Phys. J. C}\ }\textbf {\bibinfo {volume} {82}},\ \bibinfo {eid} {948} (\bibinfo {year} {2022})},\ \Eprint {http://arxiv.org/abs/2210.13750} {arXiv:2210.13750 [gr-qc]} \BibitemShut {NoStop}%
\bibitem [{\citenamefont {{Raza}}\ \emph {et~al.}(2024)\citenamefont {{Raza}}, \citenamefont {{Zubair}},\ and\ \citenamefont {{Maqsood}}}]{2024JCAP...05..047R}%
  \BibitemOpen
  \bibfield  {author} {\bibinfo {author} {\bibfnamefont {M.~A.}\ \bibnamefont {{Raza}}}, \bibinfo {author} {\bibfnamefont {M.}~\bibnamefont {{Zubair}}}, \ and\ \bibinfo {author} {\bibfnamefont {E.}~\bibnamefont {{Maqsood}}},\ }\href {\doibase 10.1088/1475-7516/2024/05/047} {\bibfield  {journal} {\bibinfo  {journal} {J. Cosmol. Astropart. Phys.}\ }\textbf {\bibinfo {volume} {2024}},\ \bibinfo {eid} {047} (\bibinfo {year} {2024})},\ \Eprint {http://arxiv.org/abs/2401.04779} {arXiv:2401.04779 [gr-qc]} \BibitemShut {NoStop}%
\bibitem [{\citenamefont {Uktamov}\ \emph {et~al.}(2025{\natexlab{a}})\citenamefont {Uktamov}, \citenamefont {Shaymatov}, \citenamefont {Ahmedov},\ and\ \citenamefont {Yuan}}]{UktamjonUktamov:2025emm}%
  \BibitemOpen
  \bibfield  {author} {\bibinfo {author} {\bibfnamefont {U.}~\bibnamefont {Uktamov}}, \bibinfo {author} {\bibfnamefont {S.}~\bibnamefont {Shaymatov}}, \bibinfo {author} {\bibfnamefont {B.}~\bibnamefont {Ahmedov}}, \ and\ \bibinfo {author} {\bibfnamefont {C.}~\bibnamefont {Yuan}},\ }\href {\doibase 10.1140/epjc/s10052-025-15171-0} {\bibfield  {journal} {\bibinfo  {journal} {Eur. Phys. J. C}\ }\textbf {\bibinfo {volume} {85}},\ \bibinfo {pages} {1432} (\bibinfo {year} {2025}{\natexlab{a}})}\BibitemShut {NoStop}%
\bibitem [{\citenamefont {{Toshmatov}}\ \emph {et~al.}(2017)\citenamefont {{Toshmatov}}, \citenamefont {{Bambi}}, \citenamefont {{Ahmedov}}, \citenamefont {{Abdujabbarov}},\ and\ \citenamefont {{Stuchl{\'\i}k}}}]{2017EPJC...77..542T}%
  \BibitemOpen
  \bibfield  {author} {\bibinfo {author} {\bibfnamefont {B.}~\bibnamefont {{Toshmatov}}}, \bibinfo {author} {\bibfnamefont {C.}~\bibnamefont {{Bambi}}}, \bibinfo {author} {\bibfnamefont {B.}~\bibnamefont {{Ahmedov}}}, \bibinfo {author} {\bibfnamefont {A.}~\bibnamefont {{Abdujabbarov}}}, \ and\ \bibinfo {author} {\bibfnamefont {Z.}~\bibnamefont {{Stuchl{\'\i}k}}},\ }\href {\doibase 10.1140/epjc/s10052-017-5112-2} {\bibfield  {journal} {\bibinfo  {journal} {Eur. Phys. J. C}\ }\textbf {\bibinfo {volume} {77}},\ \bibinfo {eid} {542} (\bibinfo {year} {2017})},\ \Eprint {http://arxiv.org/abs/1702.06855} {arXiv:1702.06855 [gr-qc]} \BibitemShut {NoStop}%
\bibitem [{\citenamefont {{Azreg-A{\"\i}nou}}(2014)}]{2014PhRvD..90f4041A}%
  \BibitemOpen
  \bibfield  {author} {\bibinfo {author} {\bibfnamefont {M.}~\bibnamefont {{Azreg-A{\"\i}nou}}},\ }\href {\doibase 10.1103/PhysRevD.90.064041} {\bibfield  {journal} {\bibinfo  {journal} {Phy.Rev.D}\ }\textbf {\bibinfo {volume} {90}},\ \bibinfo {eid} {064041} (\bibinfo {year} {2014})},\ \Eprint {http://arxiv.org/abs/1405.2569} {arXiv:1405.2569 [gr-qc]} \BibitemShut {NoStop}%
\bibitem [{\citenamefont {Azreg-A{\"\i}nou}(2014)}]{Azreg-Ainou:2014aqa}%
  \BibitemOpen
  \bibfield  {author} {\bibinfo {author} {\bibfnamefont {M.}~\bibnamefont {Azreg-A{\"\i}nou}},\ }\href {\doibase 10.1140/epjc/s10052-014-2865-8} {\bibfield  {journal} {\bibinfo  {journal} {Eur. Phys. J. C}\ }\textbf {\bibinfo {volume} {74}},\ \bibinfo {pages} {2865} (\bibinfo {year} {2014})},\ \Eprint {http://arxiv.org/abs/1401.4292} {arXiv:1401.4292 [gr-qc]} \BibitemShut {NoStop}%
\bibitem [{\citenamefont {{Azreg-A{\"\i}nou}}(2014)}]{2014PhLB..730...95A}%
  \BibitemOpen
  \bibfield  {author} {\bibinfo {author} {\bibfnamefont {M.}~\bibnamefont {{Azreg-A{\"\i}nou}}},\ }\href {\doibase 10.1016/j.physletb.2014.01.041} {\bibfield  {journal} {\bibinfo  {journal} {Physics Letters B}\ }\textbf {\bibinfo {volume} {730}},\ \bibinfo {pages} {95} (\bibinfo {year} {2014})},\ \Eprint {http://arxiv.org/abs/1401.0787} {arXiv:1401.0787 [gr-qc]} \BibitemShut {NoStop}%
\bibitem [{\citenamefont {{Newman}}\ and\ \citenamefont {{Janis}}(1965)}]{Newman}%
  \BibitemOpen
  \bibfield  {author} {\bibinfo {author} {\bibfnamefont {E.~T.}\ \bibnamefont {{Newman}}}\ and\ \bibinfo {author} {\bibfnamefont {A.~I.}\ \bibnamefont {{Janis}}},\ }\href {\doibase 10.1063/1.1704350} {\bibfield  {journal} {\bibinfo  {journal} {Journal of Mathematical Physics}\ }\textbf {\bibinfo {volume} {6}},\ \bibinfo {pages} {915} (\bibinfo {year} {1965})}\BibitemShut {NoStop}%
\bibitem [{\citenamefont {{Jusufi}}\ \emph {et~al.}(2019)\citenamefont {{Jusufi}}, \citenamefont {{Jamil}}, \citenamefont {{Salucci}}, \citenamefont {{Zhu}},\ and\ \citenamefont {{Haroon}}}]{2019PhRvD.100d4012J}%
  \BibitemOpen
  \bibfield  {author} {\bibinfo {author} {\bibfnamefont {K.}~\bibnamefont {{Jusufi}}}, \bibinfo {author} {\bibfnamefont {M.}~\bibnamefont {{Jamil}}}, \bibinfo {author} {\bibfnamefont {P.}~\bibnamefont {{Salucci}}}, \bibinfo {author} {\bibfnamefont {T.}~\bibnamefont {{Zhu}}}, \ and\ \bibinfo {author} {\bibfnamefont {S.}~\bibnamefont {{Haroon}}},\ }\href {\doibase 10.1103/PhysRevD.100.044012} {\bibfield  {journal} {\bibinfo  {journal} {Phys. Rev. D}\ }\textbf {\bibinfo {volume} {100}},\ \bibinfo {eid} {044012} (\bibinfo {year} {2019})},\ \Eprint {http://arxiv.org/abs/1905.11803} {arXiv:1905.11803 [physics.gen-ph]} \BibitemShut {NoStop}%
\bibitem [{\citenamefont {{Burinskii}}\ \emph {et~al.}(2002)\citenamefont {{Burinskii}}, \citenamefont {{Elizalde}}, \citenamefont {{Hildebrandt}},\ and\ \citenamefont {{Magli}}}]{2002PhRvD..65f4039B}%
  \BibitemOpen
  \bibfield  {author} {\bibinfo {author} {\bibfnamefont {A.}~\bibnamefont {{Burinskii}}}, \bibinfo {author} {\bibfnamefont {E.}~\bibnamefont {{Elizalde}}}, \bibinfo {author} {\bibfnamefont {S.~R.}\ \bibnamefont {{Hildebrandt}}}, \ and\ \bibinfo {author} {\bibfnamefont {G.}~\bibnamefont {{Magli}}},\ }\href {\doibase 10.1103/PhysRevD.65.064039} {\bibfield  {journal} {\bibinfo  {journal} {Phys. Rev. D.}\ }\textbf {\bibinfo {volume} {65}},\ \bibinfo {eid} {064039} (\bibinfo {year} {2002})},\ \Eprint {http://arxiv.org/abs/gr-qc/0109085} {arXiv:gr-qc/0109085 [gr-qc]} \BibitemShut {NoStop}%
\bibitem [{\citenamefont {{Khodadi}}\ \emph {et~al.}(2024)\citenamefont {{Khodadi}}, \citenamefont {{Vagnozzi}},\ and\ \citenamefont {{Firouzjaee}}}]{2024NatSR..1426932K}%
  \BibitemOpen
  \bibfield  {author} {\bibinfo {author} {\bibfnamefont {M.}~\bibnamefont {{Khodadi}}}, \bibinfo {author} {\bibfnamefont {S.}~\bibnamefont {{Vagnozzi}}}, \ and\ \bibinfo {author} {\bibfnamefont {J.~T.}\ \bibnamefont {{Firouzjaee}}},\ }\href {\doibase 10.1038/s41598-024-78264-y} {\bibfield  {journal} {\bibinfo  {journal} {Scientific Reports}\ }\textbf {\bibinfo {volume} {14}},\ \bibinfo {eid} {26932} (\bibinfo {year} {2024})},\ \Eprint {http://arxiv.org/abs/2408.03241} {arXiv:2408.03241 [gr-qc]} \BibitemShut {NoStop}%
\bibitem [{\citenamefont {Tsukamoto}\ \emph {et~al.}(2014)\citenamefont {Tsukamoto}, \citenamefont {Li},\ and\ \citenamefont {Bambi}}]{Tsukamoto:2014tja}%
  \BibitemOpen
  \bibfield  {author} {\bibinfo {author} {\bibfnamefont {N.}~\bibnamefont {Tsukamoto}}, \bibinfo {author} {\bibfnamefont {Z.}~\bibnamefont {Li}}, \ and\ \bibinfo {author} {\bibfnamefont {C.}~\bibnamefont {Bambi}},\ }\href {\doibase 10.1088/1475-7516/2014/06/043} {\bibfield  {journal} {\bibinfo  {journal} {JCAP}\ }\textbf {\bibinfo {volume} {06}},\ \bibinfo {pages} {043} (\bibinfo {year} {2014})},\ \Eprint {http://arxiv.org/abs/1403.0371} {arXiv:1403.0371 [gr-qc]} \BibitemShut {NoStop}%
\bibitem [{\citenamefont {Ono}\ \emph {et~al.}(2017)\citenamefont {Ono}, \citenamefont {Ishihara},\ and\ \citenamefont {Asada}}]{Ono:2017pie}%
  \BibitemOpen
  \bibfield  {author} {\bibinfo {author} {\bibfnamefont {T.}~\bibnamefont {Ono}}, \bibinfo {author} {\bibfnamefont {A.}~\bibnamefont {Ishihara}}, \ and\ \bibinfo {author} {\bibfnamefont {H.}~\bibnamefont {Asada}},\ }\href {\doibase 10.1103/PhysRevD.96.104037} {\bibfield  {journal} {\bibinfo  {journal} {Phys. Rev. D}\ }\textbf {\bibinfo {volume} {96}},\ \bibinfo {pages} {104037} (\bibinfo {year} {2017})},\ \Eprint {http://arxiv.org/abs/1704.05615} {arXiv:1704.05615 [gr-qc]} \BibitemShut {NoStop}%
\bibitem [{\citenamefont {{\"O}vg{\"u}n}\ \emph {et~al.}(2019)\citenamefont {{\"O}vg{\"u}n}, \citenamefont {Sakall{\i}},\ and\ \citenamefont {Saavedra}}]{Ovgun:2018fte}%
  \BibitemOpen
  \bibfield  {author} {\bibinfo {author} {\bibfnamefont {A.}~\bibnamefont {{\"O}vg{\"u}n}}, \bibinfo {author} {\bibfnamefont {{\.I}.}~\bibnamefont {Sakall{\i}}}, \ and\ \bibinfo {author} {\bibfnamefont {J.}~\bibnamefont {Saavedra}},\ }\href {\doibase 10.1016/j.aop.2019.167978} {\bibfield  {journal} {\bibinfo  {journal} {Annals Phys.}\ }\textbf {\bibinfo {volume} {411}},\ \bibinfo {pages} {167978} (\bibinfo {year} {2019})},\ \Eprint {http://arxiv.org/abs/1806.06453} {arXiv:1806.06453 [gr-qc]} \BibitemShut {NoStop}%
\bibitem [{\citenamefont {Ishihara}\ \emph {et~al.}(2016)\citenamefont {Ishihara}, \citenamefont {Suzuki}, \citenamefont {Ono}, \citenamefont {Kitamura},\ and\ \citenamefont {Asada}}]{Ishihara:2016vdc}%
  \BibitemOpen
  \bibfield  {author} {\bibinfo {author} {\bibfnamefont {A.}~\bibnamefont {Ishihara}}, \bibinfo {author} {\bibfnamefont {Y.}~\bibnamefont {Suzuki}}, \bibinfo {author} {\bibfnamefont {T.}~\bibnamefont {Ono}}, \bibinfo {author} {\bibfnamefont {T.}~\bibnamefont {Kitamura}}, \ and\ \bibinfo {author} {\bibfnamefont {H.}~\bibnamefont {Asada}},\ }\href {\doibase 10.1103/PhysRevD.94.084015} {\bibfield  {journal} {\bibinfo  {journal} {Phys. Rev. D}\ }\textbf {\bibinfo {volume} {94}},\ \bibinfo {pages} {084015} (\bibinfo {year} {2016})},\ \Eprint {http://arxiv.org/abs/1604.08308} {arXiv:1604.08308 [gr-qc]} \BibitemShut {NoStop}%
\bibitem [{\citenamefont {{Zahid}}\ \emph {et~al.}(2023)\citenamefont {{Zahid}}, \citenamefont {{Rayimbaev}}, \citenamefont {{Sarikulov}}, \citenamefont {{Khan}},\ and\ \citenamefont {{Ren}}}]{2023EPJC...83..855Z}%
  \BibitemOpen
  \bibfield  {author} {\bibinfo {author} {\bibfnamefont {M.}~\bibnamefont {{Zahid}}}, \bibinfo {author} {\bibfnamefont {J.}~\bibnamefont {{Rayimbaev}}}, \bibinfo {author} {\bibfnamefont {F.}~\bibnamefont {{Sarikulov}}}, \bibinfo {author} {\bibfnamefont {S.~U.}\ \bibnamefont {{Khan}}}, \ and\ \bibinfo {author} {\bibfnamefont {J.}~\bibnamefont {{Ren}}},\ }\href {\doibase 10.1140/epjc/s10052-023-12025-5} {\bibfield  {journal} {\bibinfo  {journal} {Eur. Phys. J. C}\ }\textbf {\bibinfo {volume} {83}},\ \bibinfo {eid} {855} (\bibinfo {year} {2023})}\BibitemShut {NoStop}%
\bibitem [{\citenamefont {{Abdujabbarov}}\ \emph {et~al.}(2015)\citenamefont {{Abdujabbarov}}, \citenamefont {{Rezzolla}},\ and\ \citenamefont {{Ahmedov}}}]{2015MNRAS.454.2423A}%
  \BibitemOpen
  \bibfield  {author} {\bibinfo {author} {\bibfnamefont {A.~A.}\ \bibnamefont {{Abdujabbarov}}}, \bibinfo {author} {\bibfnamefont {L.}~\bibnamefont {{Rezzolla}}}, \ and\ \bibinfo {author} {\bibfnamefont {B.~J.}\ \bibnamefont {{Ahmedov}}},\ }\href {\doibase 10.1093/mnras/stv2079} {\bibfield  {journal} {\bibinfo  {journal} {Mon. Not. R. Astron. Soc.}\ }\textbf {\bibinfo {volume} {454}},\ \bibinfo {pages} {2423} (\bibinfo {year} {2015})},\ \Eprint {http://arxiv.org/abs/1503.09054} {arXiv:1503.09054 [gr-qc]} \BibitemShut {NoStop}%
\bibitem [{\citenamefont {{Event Horizon Telescope Collaboration}}\ \emph {et~al.}(2019)\citenamefont {{Event Horizon Telescope Collaboration}}, \citenamefont {{Akiyama}},\ and\ \citenamefont {et~al.}}]{2019ApJ...875L...1E}%
  \BibitemOpen
  \bibfield  {author} {\bibinfo {author} {\bibnamefont {{Event Horizon Telescope Collaboration}}}, \bibinfo {author} {\bibfnamefont {K.}~\bibnamefont {{Akiyama}}}, \ and\ \bibinfo {author} {\bibnamefont {et~al.}},\ }\href {\doibase 10.3847/2041-8213/ab0ec7} {\bibfield  {journal} {\bibinfo  {journal} {Astrophys. J. Lett}\ }\textbf {\bibinfo {volume} {875}},\ \bibinfo {eid} {L1} (\bibinfo {year} {2019})},\ \Eprint {http://arxiv.org/abs/1906.11238} {arXiv:1906.11238 [astro-ph.GA]} \BibitemShut {NoStop}%
\bibitem [{\citenamefont {Abuter}\ \emph {et~al.}(2022)\citenamefont {Abuter} \emph {et~al.}}]{GRAVITY:2021xju}%
  \BibitemOpen
  \bibfield  {author} {\bibinfo {author} {\bibfnamefont {R.}~\bibnamefont {Abuter}} \emph {et~al.} (\bibinfo {collaboration} {GRAVITY}),\ }\href {\doibase 10.1051/0004-6361/202142465} {\bibfield  {journal} {\bibinfo  {journal} {Astron. Astrophys.}\ }\textbf {\bibinfo {volume} {657}},\ \bibinfo {pages} {L12} (\bibinfo {year} {2022})},\ \Eprint {http://arxiv.org/abs/2112.07478} {arXiv:2112.07478 [astro-ph.GA]} \BibitemShut {NoStop}%
\bibitem [{\citenamefont {Cautun}\ \emph {et~al.}(2020)\citenamefont {Cautun}, \citenamefont {Benitez-Llambay}, \citenamefont {Deason}, \citenamefont {Frenk}, \citenamefont {Fattahi}, \citenamefont {G{\'o}mez}, \citenamefont {Grand}, \citenamefont {Oman}, \citenamefont {Navarro},\ and\ \citenamefont {Simpson}}]{Cautun:2019eaf}%
  \BibitemOpen
  \bibfield  {author} {\bibinfo {author} {\bibfnamefont {M.}~\bibnamefont {Cautun}}, \bibinfo {author} {\bibfnamefont {A.}~\bibnamefont {Benitez-Llambay}}, \bibinfo {author} {\bibfnamefont {A.~J.}\ \bibnamefont {Deason}}, \bibinfo {author} {\bibfnamefont {C.~S.}\ \bibnamefont {Frenk}}, \bibinfo {author} {\bibfnamefont {A.}~\bibnamefont {Fattahi}}, \bibinfo {author} {\bibfnamefont {F.~A.}\ \bibnamefont {G{\'o}mez}}, \bibinfo {author} {\bibfnamefont {R.~J.~J.}\ \bibnamefont {Grand}}, \bibinfo {author} {\bibfnamefont {K.~A.}\ \bibnamefont {Oman}}, \bibinfo {author} {\bibfnamefont {J.~F.}\ \bibnamefont {Navarro}}, \ and\ \bibinfo {author} {\bibfnamefont {C.~M.}\ \bibnamefont {Simpson}},\ }\href {\doibase 10.1093/mnras/staa1017} {\bibfield  {journal} {\bibinfo  {journal} {Mon. Not. Roy. Astron. Soc.}\ }\textbf {\bibinfo {volume} {494}},\ \bibinfo {pages} {4291} (\bibinfo {year} {2020})},\ \Eprint {http://arxiv.org/abs/1911.04557} {arXiv:1911.04557 [astro-ph.GA]} \BibitemShut {NoStop}%
\bibitem [{\citenamefont {{Vincent}}\ \emph {et~al.}(2011)\citenamefont {{Vincent}}, \citenamefont {{Paumard}}, \citenamefont {{Gourgoulhon}},\ and\ \citenamefont {{Perrin}}}]{2011CQGra..28v5011V}%
  \BibitemOpen
  \bibfield  {author} {\bibinfo {author} {\bibfnamefont {F.~H.}\ \bibnamefont {{Vincent}}}, \bibinfo {author} {\bibfnamefont {T.}~\bibnamefont {{Paumard}}}, \bibinfo {author} {\bibfnamefont {E.}~\bibnamefont {{Gourgoulhon}}}, \ and\ \bibinfo {author} {\bibfnamefont {G.}~\bibnamefont {{Perrin}}},\ }\href {\doibase 10.1088/0264-9381/28/22/225011} {\bibfield  {journal} {\bibinfo  {journal} {Classical and Quantum Gravity}\ }\textbf {\bibinfo {volume} {28}},\ \bibinfo {eid} {225011} (\bibinfo {year} {2011})},\ \Eprint {http://arxiv.org/abs/1109.4769} {arXiv:1109.4769 [gr-qc]} \BibitemShut {NoStop}%
\bibitem [{\citenamefont {{Marck}}(1996)}]{1996CQGra..13..393M}%
  \BibitemOpen
  \bibfield  {author} {\bibinfo {author} {\bibfnamefont {J.-A.}\ \bibnamefont {{Marck}}},\ }\href {\doibase 10.1088/0264-9381/13/3/007} {\bibfield  {journal} {\bibinfo  {journal} {Classical and Quantum Gravity}\ }\textbf {\bibinfo {volume} {13}},\ \bibinfo {pages} {393} (\bibinfo {year} {1996})},\ \Eprint {http://arxiv.org/abs/gr-qc/9505010} {arXiv:gr-qc/9505010 [gr-qc]} \BibitemShut {NoStop}%
\bibitem [{\citenamefont {Khan}\ \emph {et~al.}(2024)\citenamefont {Khan}, \citenamefont {Uktamov}, \citenamefont {Rayimbaev}, \citenamefont {Abdujabbarov}, \citenamefont {Ibragimov},\ and\ \citenamefont {Chen}}]{Khan:2024jez}%
  \BibitemOpen
  \bibfield  {author} {\bibinfo {author} {\bibfnamefont {S.~U.}\ \bibnamefont {Khan}}, \bibinfo {author} {\bibfnamefont {U.}~\bibnamefont {Uktamov}}, \bibinfo {author} {\bibfnamefont {J.}~\bibnamefont {Rayimbaev}}, \bibinfo {author} {\bibfnamefont {A.}~\bibnamefont {Abdujabbarov}}, \bibinfo {author} {\bibfnamefont {I.}~\bibnamefont {Ibragimov}}, \ and\ \bibinfo {author} {\bibfnamefont {Z.-M.}\ \bibnamefont {Chen}},\ }\href {\doibase 10.1140/epjc/s10052-024-12567-2} {\bibfield  {journal} {\bibinfo  {journal} {Eur. Phys. J. C}\ }\textbf {\bibinfo {volume} {84}},\ \bibinfo {pages} {203} (\bibinfo {year} {2024})}\BibitemShut {NoStop}%
\bibitem [{\citenamefont {Uktamov}\ \emph {et~al.}(2025{\natexlab{b}})\citenamefont {Uktamov}, \citenamefont {Alloqulov}, \citenamefont {Shaymatov}, \citenamefont {Zhu},\ and\ \citenamefont {Ahmedov}}]{Uktamov:2024ckf}%
  \BibitemOpen
  \bibfield  {author} {\bibinfo {author} {\bibfnamefont {U.}~\bibnamefont {Uktamov}}, \bibinfo {author} {\bibfnamefont {M.}~\bibnamefont {Alloqulov}}, \bibinfo {author} {\bibfnamefont {S.}~\bibnamefont {Shaymatov}}, \bibinfo {author} {\bibfnamefont {T.}~\bibnamefont {Zhu}}, \ and\ \bibinfo {author} {\bibfnamefont {B.}~\bibnamefont {Ahmedov}},\ }\href {\doibase 10.1016/j.dark.2024.101743} {\bibfield  {journal} {\bibinfo  {journal} {Phys. Dark Univ.}\ }\textbf {\bibinfo {volume} {47}},\ \bibinfo {pages} {101743} (\bibinfo {year} {2025}{\natexlab{b}})},\ \Eprint {http://arxiv.org/abs/2412.01809} {arXiv:2412.01809 [gr-qc]} \BibitemShut {NoStop}%
\bibitem [{\citenamefont {Alloqulov}\ \emph {et~al.}(2024)\citenamefont {Alloqulov}, \citenamefont {Shaymatov}, \citenamefont {Ahmedov},\ and\ \citenamefont {Jawad}}]{1Alloqulov2024}%
  \BibitemOpen
  \bibfield  {author} {\bibinfo {author} {\bibfnamefont {M.}~\bibnamefont {Alloqulov}}, \bibinfo {author} {\bibfnamefont {S.}~\bibnamefont {Shaymatov}}, \bibinfo {author} {\bibfnamefont {B.}~\bibnamefont {Ahmedov}}, \ and\ \bibinfo {author} {\bibfnamefont {A.}~\bibnamefont {Jawad}},\ }\href {\doibase 10.1088/1674-1137/ad137f} {\bibfield  {journal} {\bibinfo  {journal} {Chinese Physics C}\ }\textbf {\bibinfo {volume} {48}} (\bibinfo {year} {2024}),\ 10.1088/1674-1137/ad137f}\BibitemShut {NoStop}%
\bibitem [{\citenamefont {Alloqulov}\ and\ \citenamefont {Shaymatov}(2024)}]{Alloqulov2024}%
  \BibitemOpen
  \bibfield  {author} {\bibinfo {author} {\bibfnamefont {M.}~\bibnamefont {Alloqulov}}\ and\ \bibinfo {author} {\bibfnamefont {S.}~\bibnamefont {Shaymatov}},\ }\href {\doibase 10.1140/epjp/s13360-024-05524-1} {\bibfield  {journal} {\bibinfo  {journal} {European Physical Journal Plus}\ }\textbf {\bibinfo {volume} {139}} (\bibinfo {year} {2024}),\ 10.1140/epjp/s13360-024-05524-1}\BibitemShut {NoStop}%
\bibitem [{\citenamefont {Alloqulov}\ \emph {et~al.}(2025{\natexlab{a}})\citenamefont {Alloqulov}, \citenamefont {Jamil}, \citenamefont {Shaymatov}, \citenamefont {Wu},\ and\ \citenamefont {Azreg-Aïnou}}]{Alloqulov2025}%
  \BibitemOpen
  \bibfield  {author} {\bibinfo {author} {\bibfnamefont {M.}~\bibnamefont {Alloqulov}}, \bibinfo {author} {\bibfnamefont {M.}~\bibnamefont {Jamil}}, \bibinfo {author} {\bibfnamefont {S.}~\bibnamefont {Shaymatov}}, \bibinfo {author} {\bibfnamefont {Q.}~\bibnamefont {Wu}}, \ and\ \bibinfo {author} {\bibfnamefont {M.}~\bibnamefont {Azreg-Aïnou}},\ }\href {\doibase 10.1016/j.jheap.2025.100424} {\bibfield  {journal} {\bibinfo  {journal} {Journal of High Energy Astrophysics}\ }\textbf {\bibinfo {volume} {48}} (\bibinfo {year} {2025}{\natexlab{a}}),\ 10.1016/j.jheap.2025.100424}\BibitemShut {NoStop}%
\bibitem [{\citenamefont {Alloqulov}\ \emph {et~al.}(2025{\natexlab{b}})\citenamefont {Alloqulov}, \citenamefont {Abdujabbarov}, \citenamefont {Ahmedov}, \citenamefont {Yuan},\ and\ \citenamefont {Zhou}}]{1Alloqulov2025}%
  \BibitemOpen
  \bibfield  {author} {\bibinfo {author} {\bibfnamefont {M.}~\bibnamefont {Alloqulov}}, \bibinfo {author} {\bibfnamefont {A.}~\bibnamefont {Abdujabbarov}}, \bibinfo {author} {\bibfnamefont {B.}~\bibnamefont {Ahmedov}}, \bibinfo {author} {\bibfnamefont {C.}~\bibnamefont {Yuan}}, \ and\ \bibinfo {author} {\bibfnamefont {C.}~\bibnamefont {Zhou}},\ }\href {\doibase 10.1140/epjc/s10052-025-14514-1} {\bibfield  {journal} {\bibinfo  {journal} {European Physical Journal C}\ }\textbf {\bibinfo {volume} {85}} (\bibinfo {year} {2025}{\natexlab{b}}),\ 10.1140/epjc/s10052-025-14514-1}\BibitemShut {NoStop}%
\end{thebibliography}%
\end{document}